\documentclass[final,3p,times]{elsarticle}

\usepackage{multirow}
\usepackage{graphicx}
\usepackage{amsmath}
\usepackage{siunitx}
\usepackage{longtable,tabularx}
\usepackage{physics}
\usepackage{amssymb}
\usepackage{physics}
\usepackage{graphicx}
\usepackage{mathrsfs}
\usepackage{color}
\usepackage{diagbox}
\usepackage{subcaption}
\usepackage{arydshln}

\makeatletter
\def\ps@pprintTitle{%
 \let\@oddhead\@empty
 \let\@evenhead\@empty
 \def\@oddfoot{}%
 \let\@evenfoot\@oddfoot}
\makeatother

\begin{document}

 \begin{frontmatter}

\title{Mesh-Free Hydrodynamic Stability}

% \tnoteref{label1}
% \tnotetext[label1]{This
% document is the results of the research project funded by the National Science Foundation Grant under No. CBET-1953999.}

% \tnotetext[label1]{Corresponding author.}

\author{Tianyi Chu}
            \ead{tic173@ucsd.edu}
\author{Oliver T. Schmidt\corref{cor1}}
            \ead{oschmidt@ucsd.edu}
            
\address{Department of Mechanical and Aerospace Engineering,  University of California San Diego, 9500 Gilman Drive, La Jolla, CA 92093, USA.}

 \cortext[cor1]{Corresponding author}

\begin{abstract}

% ------------------------

% on a range of known?

% but also to reveal (new) perspective

% validated on and provide a new perspective on well-known flow instabilities. 

A specialized mesh-free radial basis function-based finite difference (RBF-FD) discretization is used to solve the large eigenvalue problems arising in hydrodynamic stability analyses of flows in complex domains. Polyharmonic spline functions with polynomial augmentation (PHS+poly) are used to construct the discrete linearized incompressible and compressible Navier-Stokes operators on scattered nodes. Rigorous global and local eigenvalue stability studies of these global operators and their constituent RBF stencils provide a set of parameters that guarantee stability while balancing accuracy and computational efficiency. Specialized elliptical stencils to compute boundary-normal derivatives are introduced and the treatment of the pole singularity in cylindrical coordinates is discussed. The numerical framework is demonstrated and validated on a number of hydrodynamic stability methods ranging from classical linear theory of laminar flows to state-of-the-art non-modal approaches that are applicable to turbulent mean flows. The examples include linear stability, resolvent, and wavemaker analyses of cylinder flow at Reynolds numbers ranging from 47 to 180, and resolvent and wavemaker analyses of the self-similar flat-plate boundary layer at a Reynolds number as well as the turbulent mean of a high-Reynolds-number transonic jet at Mach number $0.9$. All previously-known results are found in close agreement with the literature. Finally, the resolvent-based wavemaker analyses of the Blasius boundary layer and turbulent jet flows offer new physical insight into the modal and non-modal growth in these flows.

% The comparisons of these benchmark problems with the literature allow
% us to highlight the viability, accuracy, and robustness. 

% The study furthermore encapsulates the first application of
% RA-based WM analysis to the ZPG Blasius boundary layer and turbulent jets, providing a new perspective on modal
% and non-modal growth in these flows.

% \item 

% The accuracy, viability, and robustness of the proposed mesh-free framework are highlighted by these three benchmark problems for open flows with strong instabilities.

% Our study extends beyond validating known flow cases by introducing a pioneering application of RA-based WM analysis to the ZPG Blasius boundary layer and turbulent jets, providing fresh physical insights into the modal and non-modal growth in these flows.

% Radial basis function-based finite differences (RBF-FD) are used to develop a high-order mesh-free hydrodynamic stability analysis tool for complex geometries. Polyharmonic spline RBFs with polynomial augmentation (PHS+poly) are used to construct the discrete linearized Navier-Stokes and resolvent operators on arbitrarily scattered nodes. The PHS+poly discretization is shown to yield accurate, stable, and computationally efficient discretizations of the large matrix problems arising in two-dimensional hydrodynamic stability analysis.

% The development of a high-order, mesh-free framework for global hydrodynamic stability analysis is present. Efficient construction of global Jacobians on scattered nodes based on the PHS+poly RBF-FD discretizations is demonstrated.

\end{abstract}

\begin{keyword}

RBF-FD  \sep polyharmonic splines \sep polynomial augmentation \sep linear stability theory\sep Navier-Stokes \sep wavemaker

%% keywords here, in the form: keyword \sep keyword

%% PACS codes here, in the form: \PACS code \sep code

%% MSC codes here, in the form: \MSC code \sep code
%% or \MSC[2008] code \sep code (2000 is the default)

\end{keyword}

 \end{frontmatter}

 \section{Introduction}

% \Red{[classic linear stability (local/global)]}

    Flow instabilities and large-scale coherent structures are ubiquitous phenomena in fluid mechanics that have been the focus of extensive research. 
Linear stability (LST) analysis is specifically designed to investigate the growth of small perturbations exclusively around laminar base flows, which are 
the steady-state solution to the Navier-Stokes equations.
One-dimensional LST analysis was widely used in the past century, e.g., \citep{betchov1963stability,crighton1976stability,ho1984perturbed, mattingly1972stability}.
\citet{eriksson1985computer} and \citet{tuckerman1985formation} were the pioneers in conducting LST analysis in two-dimensional (2D).
Subsequently, \citet{jackson1987finite} and \citet {zebib1987stability} examined the 2D nature of vortex shedding in the wakes of bluff bodies. Readers are referred to \citet{huerre1990local} and \citet{theofilis2011global} for comprehensive reviews of the concept of 2D LST modes. The implementation of the 2D LST framework has enabled improved identification of flow instability in non-parallel flows, including
cylinder wakes \cite{giannetti2007structural, marquet2008sensitivity, noack1994global}, aerofoil wakes \citep{edstrand2018parallel, woodley1997global}, boundary layers \citep{aakervik2008global,ehrenstein2005two},
and jets in cross-flow \citep{bagheri2009global,peplinski2015global,regan2017global}. 
Studies have demonstrated that an open flow can possess marginal stability despite exhibiting local convective instability.
LST analysis around a steady laminar base flow, by its inherent nature, is not applicable to predict finite-amplitude flow instabilities arising from nonlinear interactions. 
The use of mean flow for LST analysis, despite violating the basic assumption of linear theory, has been used successfully to predict coherent flow features in diverse types of flows, including cylinder wakes \citep{barkley2006linear, pier2002frequency},
open cavity flows \citep{sipp2012open,sipp2007global},
mixing layers \citep{gaster1985large,monkewitz1988subharmonic,weisbrot1988coherent},
and turbulent or transitional jets \citep{gudmundsson2011instability,oberleithner2014mean,schmidt2017wavepackets}.
     Theoretical conditions required for the validity of mean flow stability analysis have been explored by \citet{beneddine2016conditions}.
Although beyond the scope of this work, it is noteworthy that the weakly-nonlinear extension of LST analysis has been successfully employed for studying the dynamics of non-parallel flows \citep{chomaz2005global,sipp2007global}.
The LST-based semi-empirical $e^N$ method \citep{van1956suggested,smith1956transition} has succeeded in transition prediction for certain flows such as boundary layers.
However, the prediction of disturbance behavior in more complex scenarios, such as crossflows or bypass transitions, falls outside the scope of LST theory.

% Despite the success semi-empirical $e^N$ method \citep{van1956suggested,smith1956transition}
% has succeeded in transition prediction for certain flows, such as boundary layers, 

% However, LST theory encounters limitations for accurately predicting the behavior of disturbances in more complex scenarios, including crossflows or bypass transitions.

% LST theory has limitations in accurately predicting the behavior of disturbances for general flows.

% Nonlinear effects such as transient growth and non-normality can lead to significant deviations from the linear predictions. 

% In addition, non-modal approaches have also demonstrated success in receptivity analysis \citep{berlin1999nonlinear,luchini2000reynolds,moarref2010controlling} and the prediction of bypass transition \citep{andersson1999optimal,jovanovic2005componentwise}. 

% As such, non-modal analysis techniques are essential to account for the evolution of finite-amplitude perturbations and provide a more accurate understanding of the mechanisms driving the onset of turbulence and the growth of coherent structures, see, e.g., \citep{reddy1993pseudospectra,schmid2007nonmodal,schmid2002stability}.

% \Red{[Resolvent analysis]}

Despite these limitations, resolvent, or input-output analysis, has recently emerged as a linear tool for accurately predicting large-scale coherent structures in fully turbulent flows. Resolvent analysis (RA) originally stems from the studies of transient growth \citep{farrell1993stochastic,reddy1993energy, reddy1993pseudospectra,trefethen1993hydrodynamic} and seeks the optimal pairs of inputs and corresponding outputs through the linearized system that maximizes the energy gain. Within the laminar regime, RA has been utilized to investigate the linear response to external body forces or perturbations for channel flows \citep{jovanovic2005componentwise,moarref2012model,rosenberg2019computing}, boundary layers \citep{alizard2022stochastic,bonne2019analysis,bugeat20193d,monokrousos2010global,pfister2022global,sipp2013characterization,towne2022efficient}, and jets \citep{garnaud2013preferred,jeun2016input,semeraro2016modeling}.
    In contrast to classical LST analysis, the input-output perspective offers a mathematically rigorous framework for studying turbulent mean-flows by identifying the forcing as the Reynolds stresses in the perturbation-interaction terms in the Reynolds-decomposed Navier–Stokes equations \citep{mckeon2010critical,sharma2013coherent}.
    Applications include near-wall flows \citep{hwang2010amplification, aakervik2008global,sharma2013coherent}, boundary layers \citep{bae2020resolvent, jacobi2011dynamic,rigas2021nonlinear}, incompressible \citep{garnaud2013preferred,lesshafft2019resolvent} or compressible jets \citep{jeun2016input, schmidt2018spectral, towne2018spectral},
and airfoil wakes \citep{ribeiro2020randomized, yeh2019resolvent}. The validation of fundamental relationships between RA and other modal decomposition techniques was facilitated by \citet{towne2018spectral}, establishing RA as a well-suited tool for turbulence modeling \citep{pickering2020lift, pickering2021optimal}.
Both classical LST and RA require, in their most basic form, the construction of large matrices that have to be decomposed into their singular- or eigen-components.
The construction and decomposition of these matrices are particularly challenging for flows in complex geometries. 
Previous studies \citep{dwivedi2022oblique, ribeiro2022wing, sidharth2018onset, yeh2019resolvent} have leveraged the flexibility of the finite-volume (FV) methods
on unstructured meshes.
However, a downside of unstructured FV methods is that the accuracy is usually restricted to 2nd-order.
Alternatively, the utilization of finite-element (FE) methods for spatial discretization provides high-order accuracy for flow instability analysis \citep{marquet2008sensitivity,marquet2009direct,rigas2021nonlinear,sipp2007global}.
The commonly employed weak formulation of governing equations in FE methods raises concerns regarding stability and convergence.
Finite-element (FE) methods offer the same flexibility, and the FreeFEM+ toolbox by \citet{hecht2012new} has been employed in a number of recent studies \citep{marquet2008sensitivity,marquet2009direct,rigas2021nonlinear,sipp2007global}.
Matrix-free methods, such as the time-stepper \citep{bagheri2009matrix, bagheri2009global, tuckerman2000bifurcation} and other related techniques \citep{de2012efficient, martini2021efficient, monokrousos2010global, ohmichi2021matrix}, provide an alternative approach where the decomposition of large matrix operations can be completely circumvented.
Iterative Krylov subspace methods are commonly employed to obtain a partial eigendecomposition. However, their major limitation lies in their capacity to extract only a limited portion of the spectra at a time.
Randomized approaches have also been explored as potential solutions to decrease the computational cost of the singular- or eigendecomposition of large stability matrices \citep{moarref2014low, ribeiro2020randomized}.
In this study, we demonstrate the capability and accuracy of radial basis functions (RBF) in effectively addressing the aforementioned challenges.

% showing that resolvent response modes are equivalent to spectral proper orthogonal decomposition (SPOD) modes under white noise excitation.

% \Red{[PHS+poly RBF-FD]}

The use of RBF-based methods provides high flexibility in meshing complex geometries, allowing for local grid refinement and local adjustments to the order of accuracy of the discretization. The RBF methodology is rooted in scattered data interpolation, first introduced by \citet{hardy1971multiquadric}, and offers a systematic means of approximating multivariate functions on arbitrarily scattered nodes.
By generalizing the classical finite-difference (FD) methods, the so-called RBF-FD methods enable the approximation of spatial derivatives to arbitrary node layouts based on RBFs \citep{tolstykh2000using}.
In fluid flow problems, Gaussian (GA), multiquadric (MQ), and inverse multiquadric (IMQ) are among the most commonly used radial basis functions (RBFs).
Refer to \citet{fornberg2015solving} for a detailed overview.
The computation of these RBFs requires specifying a shape parameter, a free parameter that can significantly impact both the numerical stability and accuracy.
Recently, a new class of RBF-FD approximations has emerged that utilizes polyharmonic splines (PHS) augmented with polynomials (PHS+poly).
This technique was first introduced by \citet{flyer2016enhancing} and has demonstrated great potential in achieving high-order accuracy while avoiding the difficulties associated with the shape parameter selection.
\citet{bayona2017role} and \citet{flyer2016role} highlighted the advantages of local PHS+poly-generated RBF-FD stencils in achieving high-order accuracy for derivative approximations and solving elliptic partial differential equations (PDEs), while also eliminating stagnation errors under node refinement.
The feasibility of implementing a larger stencil near domain boundaries to avoid the Runge phenomenon was analytically verified by \citet{bayona2019insight} using RBFs in a closed-form. 
Additional numerical examples in 2-D and 3-D were presented in \citet{bayona2019role} to demonstrate the effectiveness of the PHS+poly approach.
The accuracy and stability of the PHS+poly RBF-FD method depend on the combination of the stencil size, PHS exponent, and the degree of polynomials.
Several studies, including \citet{chu2023rbf,le2023guidelines,shankar2018hyperviscosity}, have investigated and discussed the optimal combinations of these parameters to achieve the `spectral-like' accuracy of higher-order Pad\'e-type compact FDs \citep{lele1992compact}.
In addition, other approaches have been explored to improve the computational accuracy and efficiency of PHS+poly RBF-FDs, including overlapping stencils \citep{shankar2017overlapped,shankar2018hyperviscosity}, the partition of unity (PU) method \citep{mirzaei2021direct}, and oversampling \citep{tominec2021least}.
Comparative analyses have been carried out on various higher-order mesh-free discretizations, consistently demonstrating that PHS+poly-based RBF-FDs outperform other methods in accuracy, robustness, and computational efficiency, e.g., \citep{bayona2019comparison, flyer2016role, gunderman2020transport, orucc2022strong, santos2018comparing}.
Recent successful applications of PHS+poly RBF-FDs in flow simulations on scattered nodes \citep{shahane2021high,shahane2022consistency, shahane2023semi, unnikrishnan2022shear} highlight the potential of RBF methods for investigating flow instabilities in complex geometries.
In our recent work \citep{chu2023rbf}, we introduced a staggered-node RBF-FD-based semi-implicit fractional-step solver for the incompressible Navier-Stokes equations, which will be utilized to compute the base state in the present study.

% Recently, RBF-FD approximations based on polyharmonic splines (PHS) that are augmented with polynomials (PHS+poly), as first introduced by \citet{flyer2016enhancing}, have revealed their potential to achieve high-order accuracy without handling such a parameter.
% In particular, \citet{flyer2016role} showed that local PHS+poly generated RBF-FD stencils provide high-order accuracy for derivative approximations and eliminate stagnation errors under node refinement.

% \citet{bayona2017role} utilized the PHS+poly RBF-FDs to solve elliptic partial differential equations (PDEs) and implemented a larger stencil near domain boundaries to avoid the Runge phenomenon. The feasibility of this approach was later analytically verified by \citet{bayona2019insight} using RBFs in a closed-form. 
% Additional numerical examples in 2-D and 3-D were presented in \citet{bayona2019role} to demonstrate the effectiveness of the PHS+poly approach.

% \citet{bayona2017role} used PHS+poly to solve elliptic PDEs and relied on a larger stencil near domain boundaries to avoid the Runge phenomenon. 
% The feasibility of this strategy was later analytically confirmed by \citet{bayona2019insight} using RBFs in a closed-form. 
% Numerical demonstrations for 2-D and 3-D examples were presented in \citet{bayona2019role}. 

% \Red{[main contribution]}

This work introduces a high-order, mesh-free framework for hydrodynamic stability analysis based on the PHS+poly RBF-FD discretizations.
The efficient construction of global Jacobians on scattered nodes using RBF-based methods is demonstrated.
The trade-off between computational efficiency and accuracy is addressed, and best practices are provided.
Furthermore, we discuss the practical treatment of boundary conditions, particularly the boundary-normal derivatives and the pole singularity in cylindrical coordinates.
The numerical framework is demonstrated on various hydrodynamic stability theoretical methods and flows. Refer to table \ref{flow_summary} for an overview.
In particular, LST, RA, and wavemaker (WM) analyses are shown for the canonical cylinder flow at Reynolds numbers ranging from 47 to 180,
RA and WM analyses for a laminar zero-pressure-gradient (ZPG) flat-plate Blasius boundary layer at a Reynolds number of $6\times10^{5}$,
RA and WM analyses for a turbulent transonic jet at Mach number $0.9$ and a Reynolds number of approximately $10^6$.
The identified flow phenomena include vortex shedding in cylinder wakes \citep{barkley2006linear,pier2002frequency}, the Orr and Tollmien-Schlichting waves in boundary layers \citep{aakervik2008global,brandt2011effect,monokrousos2010global,sipp2013characterization}, and Orr and Kelvin-Helmholtz wavepackets \citep{cavalieri2013wavepackets,garnaud2013preferred,lesshafft2019resolvent,schmidt2018spectral,suzuki2006instability} as well as trapped acoustic waves \citep{schmidt2017wavepackets,towne2017acoustic} in turbulent jets.
The comparisons of these benchmark problems with the literature allow us to highlight the viability, accuracy, and robustness.
The study furthermore encapsulates the first application of RA-based WM analysis to the ZPG Blasius boundary layer and turbulent jets, providing a new perspective on modal and non-modal growth in these flows.

% In the turbulent regime, the method has successfully identified both 
% Orr and Kelvin-Helmholtz wavepackets \citep{cavalieri2013wavepackets,garnaud2013preferred,lesshafft2019resolvent,schmidt2018spectral,suzuki2006instability}, as well as trapped acoustic waves \citep{schmidt2017wavepackets,towne2017acoustic}, in a transonic jet.

% \Red{[Organizations]}

This paper is organized as follows:
\S \ref{RBF} introduces the PHS+poly RBF-FD method,
\S \ref{meshfree_stability} presents the mesh-free hydrodynamic stability analysis, and \S \ref{numerical} discusses the numerical implementations. The performance of the mesh-free framework is demonstrated in \S \ref{applications}. Finally, \S \ref{conclusion} concludes and summarizes the paper.

% \newpage\clearpage
 
\section{Radial basis function-based finite differences (RBF-FD)}  \label{RBF}

The goal of the RBF-FD method is to compute the discrete representation of any linear differentiation operator $\mathcal{D}$ at a given location $x_0$ 
as a linear combination of the function values $g(\vb*{x}_j)$, such that
\begin{equation}
   \mathcal{D} g(\vb*{x}_0)=\sum_{j=1}^n w_{j} g(\vb*{x}_j). \label{Lf} 
\end{equation}
To obtain the unknown differentiation weights $w_j$, we use the RBF interpolant,
\begin{equation}
    s(\vb*{x})=\sum_{j=1}^n \gamma_j \phi(\|\vb*{x}-\vb*{x}_j\|), \label{rbf_intp}
\end{equation}
to approximate the given function $g(\vb*{x})$ by satisfying 
\begin{equation}
    s(\vb*{x}_j)=g(\vb*{x}_j), \qquad j= 1,2,\cdots n. \label{rbf_intp_cond}
\end{equation}
Here, $\phi(r)$ is a smooth radial function, $\{\vb*{x}_j\}_{j=1}^n$ is a set of scattered nodes, and $\|\cdot\|$ denotes the standard Euclidean norm. Combining equations (\ref{Lf}-\ref{rbf_intp_cond}) leads to the linear system
\begin{equation}
   \underbrace{\mqty[\phi(\|\vb*{x}_1-\vb*{x}_1\|) & \phi(\|\vb*{x}_1-\vb*{x}_2\|) & \cdots & \phi(\|\vb*{x}_1-\vb*{x}_n\|)\\
    \phi(\|\vb*{x}_2-\vb*{x}_1\|) & \phi(\|\vb*{x}_2-\vb*{x}_2\|) & \cdots & \phi(\|\vb*{x}_2-\vb*{x}_n\|)\\
    \vdots & \vdots &  & \vdots\\
       \phi(\|\vb*{x}_n-\vb*{x}_1\|) & \phi(\|\vb*{x}_n-\vb*{x}_2\|) & \cdots & \phi(\|\vb*{x}_n-\vb*{x}_n\|)
    ]}_{\vb*{A}} \mqty[ w_1 \\ w_2 \\ \vdots \\ w_n]=\mqty[ \eval{\mathcal{D}\phi(\|\vb*{x}-\vb*{x}_1\|)}_{\vb*{x}=\vb*{x}_0} \\
   \eval{\mathcal{D}\phi(\|\vb*{x}-\vb*{x}_2\|)}_{\vb*{x}=\vb*{x}_0}  \\
   \vdots \\
   \eval{\mathcal{D}\phi(\|\vb*{x}-\vb*{x}_n\|)}_{\vb*{x}=\vb*{x}_0}], \label{w_local}
\end{equation}
which can be solved directly to obtain the weight vector $\vb*{w}=[w_1,\cdots, w_n]^T$.
An implicit assumption for equation (\ref{w_local}) is that the derivative of the basis function, $\mathcal{D}\phi$, is continuous.

Polynomial augmentation is commonly applied to the RBF-FD method to enforce consistency with Taylor expansion-based FD approximations \cite{flyer2016enhancing,fornberg2015solving, fornberg2011stabilization,larsson2013stable,wright2006scattered}. The two-dimensional augmented RBF-FD method can be expressed as
\begin{equation}
   \mathcal{D}g(\vb*{x}_0)=\sum_{j=1}^n w_{j} g(\vb*{x}_j) + \sum_{i=1}^{(q+1)(q+2)/2} c_i P_i(\vb*{x}_0), \label{Lf_PHS} 
\end{equation}
where $P_i(\vb*{x})$ are multivariate polynomials up to degree $q$.
To match the local Taylor series, additional constraints for the differentiation weights,  
\begin{align}\label{rbf_poly_constrain}
    \sum_{j=1}^n w_j P_i(\vb*{x}_j) =  \mathcal{D} P_i(\vb*{x}_0) \qquad \text{for} \,\, 1\leq i\leq \frac{(q+1)(q+2)}{2},
\end{align}
are included in the computation. These constraints, also referred to as the vanishing momentum conditions \cite{iske2003approximation}, ensure that the RBF approximations locally reproduce polynomial behavior up to degree $q$ \cite{flyer2016role} and decay in the far-field \cite{fornberg2002observations}.
A more general and compact representation of equations (\ref{Lf_PHS}-\ref{rbf_poly_constrain}) are
\begin{equation} \label{rbf_poly_compact}
   \underbrace{\left[
\begin{array}{c  c }
    \vb*{A} &  \vb*{P} \\ 
   \vb*{P}^{\small{T}} & \vb*{0}
   \end{array}\right]}_{\vb*{A}_\text{aug}}
   \mqty[\vb*{w}\\ \vb*{c}]=\mqty[{\mathcal{D}\vb*{\phi}}\\
   {\mathcal{D}\vb*{P}}], 
\end{equation}
where $\vb*{A}$ represents the same interpolation matrix as defined in equation (\ref{w_local}). 
In equation (\ref{Lf}), only the weight vector $\vb*{w}$ is used to approximate the differentiation operator $\mathcal{D}$. Motivated by recent studies (see, e.g., \cite{bayona2017role,bayona2019role,flyer2016enhancing,flyer2016role}), we use polyharmonic splines (PHS),
\begin{align}
    \phi(r) =r^m,
\end{align}
 as the basis functions, where $m$ is an odd positive integer. The accuracy and stability of this PHS+poly RBF-FD method depend on the combination of the stencil size, PHS exponent, and the degree of polynomials. The selection of these parameters will be discussed later in \S\ref{parameter_selection}.

 % Later, we will identify the optimal combination, $(n,m,q)$, that is most well-suited for hydrodynamic stability analysis.

RBF-FD methods provide a mesh-free approach for discretizing the domain without the need for a pre-defined mesh or explicit node connectivity. 
This feature enables the investigation of flow instabilities for complex geometries, thereby offering advantages over traditional discretizations that rely on structured meshes. The subsequent sections of this paper focus on the construction of the discrete linearized Navier-Stokes (LNS) and then the resolvent operators using PHS+poly RBF-FDs.

\section{Mesh-free hydrodynamic stability analysis }\label{meshfree_stability}

% Orr-Sommerfeld

\subsection{Governing equations}

The motion of a general incompressible Newtonian fluid is governed by the Navier-Stokes equations,
\begin{subequations}\label{NS_eqn}
\begin{align}
    \pdv{\vb*{u}}{t}+ \left(\vb*{u}\cdot{\nabla}\right)\vb*{u} &= -{\nabla}  p +\mathrm{Re}^{-1} {\nabla} ^2 \vb*{u}, \\
    {\nabla}\cdot \vb*{u} &=0.
\end{align}
\end{subequations}
Here, all variables are nondimensionalized by the velocity scale $\vb*{U}_{\infty}$ and the length scale $L$, and $\mathrm{Re}$ denotes the Reynolds number.

In the case of laminar flows, we can decompose the flow state around the steady-state solution of the Navier-Stokes equations (\ref{NS_eqn}) as 
\begin{align}
     \vb*{u}=\vb*{U}+ \vb*{u}', \quad p={P}+  p',
\end{align}
where $(\vb*{U},\, P)$ represents the base flow that satisfies
\begin{subequations} \label{base_flow}
\begin{align}
 \left(\vb*{U}\cdot{\nabla}\right)\vb*{U} &= -{\nabla}  P +\mathrm{Re}^{-1} {\nabla} ^2 \vb*{U}, \\
    {\nabla}\cdot \vb*{U} &=0,
\end{align}
\end{subequations}
and $(\cdot)'$ denotes the small fluctuating components.
In turbulent flows, we take the Reynolds decomposition of the flow state into the temporal mean, $\overline{\left(
\cdot\right)}$, and fluctuating components, given by
\begin{align}
     \vb*{u}=\overline{\vb*{u}}+ \vb*{u}', \quad p=\overline{p}+  p'.
\end{align}
By generalizing the notation of the base state as $(\vb*{u}_0,\,p_0)$, the resulting governing equations for the fluctuations take the form of
\begin{subequations}\label{NS_eqn_fluc} 
\begin{align}
    \pdv{\vb*{u}'}{t}+\left({\vb*{u}_0}\cdot{\nabla}\right) \vb*{u}'+ \left(\vb*{u}'\cdot{\nabla} \right) {\vb*{u}_0} &= -{\nabla}  p' +\mathrm{Re}^{-1} {\nabla} ^2 \vb*{u}' + \vb*{f}', \label{NS_eqn_fluc_1} \\
    {\nabla}\cdot \vb*{u}' &=0. \label{NS_eqn_fluc_2}
\end{align}
\end{subequations}
Here, the term $\vb*{f}'$ represents the remaining nonlinear interactions between the fluctuation components. Table \ref{summary_decomposition} summarizes the two aforementioned decompositions.

\begin{table}[!htb]
      \centering
      \renewcommand{\arraystretch}{1.5}
\begin{tabular}{|c|c|c|c|c|}
\hline
Base state   & Description & Notation &  Obtained from  & Remaning forcing $\vb*{f}'$   \\ \hline \hline
\multirow{2}{*}{$(\vb*{u}_0,\,p_0)$} & Base flow   & $(\vb*{U},\, P)$                    & equation (\ref{base_flow}) & $- \left(\vb*{u}'\cdot{\nabla}\right) \vb*{u}'$   \\ \cline{2-5} 
  & Mean flow   & $(\overline{\vb*{u}},\overline{p})$ & long-time average    & $- \left(\vb*{u}'\cdot{\nabla}\right) \vb*{u}'+\overline{\left(\vb*{u}'\cdot{\nabla}\right) \vb*{u}'}$ \\ \hline
\end{tabular}
        \caption{Summary of flow state decompositions.}\label{summary_decomposition}
\end{table}

% \begin{align}
%      \vb*{u}=\vb*{u}_0+ \vb*{u}', \quad p={p}_0+  p',\label{flow_d}
% \end{align}

Equation (\ref{NS_eqn_fluc})
 can be written compactly in terms of the fluctuating state, $\vb*{q}'= [u',\,v',\,p']^T$, as
\begin{align}
   \mathcal{P}\mathcal{P}^T \left(\pdv{}{t}\vb*{q}'\right)  &= \mathcal{L} \vb*{q}' + \mathcal{P}\vb*{f}', \label{NS_eqn_fluc_compact} 
\end{align}
 where $\mathcal{P}$ is the prolongation operator that extends the velocity vector $[u,\,v]^T$ into $[u,\,v,\, 0]^T$, and its transpose is the restriction operator
that extracts the velocity vector from the extended state vector \citep{sipp2013characterization}.
The incompressible linearized Navier-Stokes (LNS) operator takes the form of
\begin{align}\label{N_S}
    \mathcal{L} \equiv \mqty*(-\left({\vb*{u}_0}\cdot{\nabla}\right)()-\left[\left(\right)\cdot{\nabla}\right]{\vb*{u}_0}+\mathrm{Re}^{-1}{\nabla}^2 & -{\nabla}  \\
    {\nabla}\cdot \left(\right) & 0).
\end{align}
The governing equations for compressible flows will be discussed in \S \ref{jet} and \ref{Governing_compressible}.
Beyond the linear dynamics, the remaining forcing $\vb*{f}'$ in equation (\ref{NS_eqn_fluc_compact}) comprises products of fluctuating quantities, as outlined in table \ref{summary_decomposition}.
These terms will be either neglected or modeled.

Classical (temporal) linear stability (LST) analysis investigates fluctuations with complex frequency $\lambda=\lambda_r+\mathrm{i} \lambda_i$, where $\lambda_r$ is the exponential growth rate and $\lambda_i$ the oscillation frequency. 
The fluctuations are assumed to be infinitesimally small, and the forcing term, $\vb*{f}'$, is, therefore, negligible at $O(1)$, see, e.g., \citet{schmid2002stability}. 
 Substituting perturbations of the form $ \vb*{q}'(\vb*{x},t) =\tilde{\vb*{q}}(\vb*{x})\mathrm{e}^{\lambda t}$ into the governing equations (\ref{NS_eqn_fluc}) 
  yields the LST equation, 
 \begin{align}\label{global_instability}
     \lambda \mathcal{P}\mathcal{P}^T \tilde{\vb*{q}}  = \mathcal{L}\tilde{\vb*{q}}.
 \end{align}
Equation (\ref{global_instability}) is a generalized eigenvalue problem, and the eigenvector associated with the largest growth rate ought to predict the dominant flow instability mechanism.

The nonlinear interactions in equation (\ref{NS_eqn_fluc_compact}) are no longer negligible for general cases of finite amplitude fluctuations.
Within the resolvent framework of turbulent flows, the nonlinear interactions, along with the background turbulence, can be interpreted as external forcing, $\vb*{f}'$, to the otherwise linear dynamics.
This interpretation was first proposed by \citet{mckeon2010critical}.
By assuming a normal mode form for the fluctuating components, $ [\vb*{q}',\,\vb*{f}'](\vb*{x},t) =    [\hat{\vb*{q}},\,\vb*{\hat{f}}](\vb*{x})\mathrm{e}^{\mathrm{i}\omega t} +c.c.$, where $\omega $ is the angular frequency, or equivalently by taking the Fourier transform, we obtain the linear time-invariant (LTI) representation of the governing equation (\ref{NS_eqn_fluc}) in the frequency domain, 
 \begin{subequations}  \label{Navier_resolvent}
   \begin{align}
    \left(\mathrm{i}\omega\mathcal{P}\mathcal{P}^T-\mathcal{L}\right) \hat{\vb*{q}}  &= \mathcal{P}\left( \mathcal{B}\vb*{\hat{f}} \right), \label{Navier_frequency}\\
    \vb*{\hat{u}}  &=\mathcal{P}^T  \left(\mathcal{C} \vb*{\hat{q}}\right). \label{Navier_output}
\end{align}
 \end{subequations}
The linear operators $\mathcal{B}$ and $\mathcal{C}$ are used to select spatial regions of particular interest.
We write equations (\ref{Navier_resolvent}) in a compact form as
\begin{align}\label{input_output}
    \hat{\vb*{u}}=\mathcal{H}(\omega)\vb*{ \hat{f}},
\end{align}
where $\mathcal{H}(\omega)=\mathcal{P}^T\mathcal{C}\left(\mathrm{i}\omega\mathcal{P}\mathcal{P}^T-\mathcal{L} \right)^{-1}\mathcal{P}\mathcal{B}$ is known as the resolvent operator. The numerical discretization of the global operator, $\mathcal{L}$,
is at the core of both LST and resolvent analyses.
In this work, the PHS+poly RBF-FDs are utilized to construct differentiation matrices and global Jacobians on scattered nodes, taking advantage of their flexibility and accuracy in numerical discretization.

\subsection{Global Jacobians and mesh-free stability analysis}

For differentiation operators $\mathcal{D}=\frac{\partial^{(\alpha+\beta)} }{\partial x_1^{\alpha} \partial x_2^{\beta} }$ with $\alpha+\beta\leq 2$, we seek differentiation matrices $\vb*{D}_{x_1^{\alpha}x_2^{\beta}}$ that satisfy
\begin{align}
   &\underbrace{\mqty[w_{11} & w_{12} & \cdots & w_{1N}\\
    w_{21} & w_{2 2} & \cdots & w_{2 N}\\
    \vdots & \vdots &  & \vdots\\
       w_{N 1} & w_{N 2} & \cdots & w_{N N}
    ]}_{ \vb*{D}_{x_1^{\alpha}x_2^{\beta}} } \mqty[ g(\vb*{x}_1) \\ g(\vb*{x}_2) \\ \vdots \\ g(\vb*{x}_N)]=\mqty[ \frac{\partial^{(\alpha+\beta)} }{\partial x_1^{\alpha} \partial x_2^{\beta} } g(\vb*{x}_1) \\ \frac{\partial^{(\alpha+\beta)} }{\partial x_1^{\alpha} \partial x_2^{\beta} } g(\vb*{x}_2) \\ \vdots \\ \frac{\partial^{(\alpha+\beta)} }{\partial x_1^{\alpha} \partial x_2^{\beta} } g(\vb*{x}_N)].
\end{align}
Here, $(x_1,x_2)$ represents the coordinate system, and $\{\vb*{x}_i\}_{i=1}^N$ represents the global computational grid.
The $j$th row of the sparse matrix $\vb*{D}$ contains the $n\ll N$ weights that approximate the derivative at node $\vb*{x}_j$. The matrix hence has $N\times n$ nonzero elements. 
In a slight change of notation, to enhance readability, we now denote by $\vb*{u}=\vb*{u}(\vb*{x})$ and $\vb*{v}=\vb*{v}(\vb*{x})$ the global velocity fields, and by $\vb*{p}=\vb*{p}(\vb*{x})$ the global pressure field in the computational domain $\Omega$.

Upon the use of these global RBF-FD-based differentiation matrices, we assemble the discrete global LNS operator from equation (\ref{N_S}), taking the two-dimensional Cartesian coordinate system as an example, as
\begin{align}\label{LNS_incomp_RBF}
    \vb*{L}  =  \mqty*(\vb*{S}-\text{diag}\left(\vb*{D}_x {\vb*{u}}_0 \right)+\mathrm{Re}^{-1}\left(\vb*{D}_{xx}+\vb*{D}_{yy}\right) & -\text{diag}\left(\vb*{D}_y {\vb*{u}}_0 \right) & -\vb*{D}_x  \\
    -\text{diag}\left(\vb*{D}_x {\vb*{v}}_0 \right)& \vb*{S}-\text{diag}\left(\vb*{D}_y {\vb*{v}}_0 \right)+\mathrm{Re}^{-1}\left(\vb*{D}_{xx}+\vb*{D}_{yy}\right)  & -\vb*{D}_y\\
    \vb*{D}_x & \vb*{D}_y & \vb*{0}),
\end{align}
where $\vb*{S} = -\left({\vb*{u}_0} \circ\vb*{D}_x + {\vb*{v}_0} \circ\vb*{D}_y\right)$, and $\circ$ denotes the Hadamard product.

For a weighted inner product $\left<{\vb*{q}}_1,{\vb*{q}}_2\right> =  {\vb*{q}}_2^*\vb*{W} {\vb*{q}}_1$ that accounts for the non-uniformly distributed nodes, the discrete LST and its adjoint eigenvalue problems take the form of
 \begin{subequations} \label{global_instability_discretized}
 \begin{align}
     \lambda \vb*{P}\vb*{P}^T \tilde{\vb*{q}}  &= \vb*{L}  \tilde{\vb*{q}} , \\ 
          \lambda^+ \vb*{P}\vb*{P}^T \tilde{\vb*{q}}^+  &=\vb*{L}^+ \tilde{\vb*{q}}^+ ,
 \end{align}
  \end{subequations} 
  respectively, where 
    \begin{align}\label{LNS_incomp_RBF_adjoint}
    \vb*{L}^+  = \vb*{W}^{-1} \vb*{L}^H \vb*{W}
\end{align} 
is the discrete adjoint LNS operator. 
Here, the restriction matrix for two-dimensional problems takes the form of $\vb*{P}^T=\mqty[\vb*{I} & \vb*{0} &\vb*{0}\\ \vb*{0} & \vb*{I} &\vb*{0}\\]$, and $\left(\cdot\right)^H$ denotes the Hermitian transpose. The eigenvectors of these two generalized eigenvalue problems, $\tilde{\vb*{q}}  = [\tilde{\vb*{u}},\tilde{\vb*{v}},\tilde{\vb*{p}} ]^T$ and $\tilde{\vb*{q}}^+ = [\tilde{\vb*{u}}^+,\tilde{\vb*{v}}^+,\tilde{\vb*{p}}^+ ]^T$, are referred to as the LST and adjoint modes, respectively.

The wavemaker (WM), introduced by \citet{giannetti2007structural}, identifies the flow region with the strongest localized feedback, where the dominant instability mechanisms act.
Based on the leading LST and adjoint modes, the WM is locally defined as
\begin{align}\label{structral_sens}
    \zeta_{\text{LST}}(x_1,x_2) = \frac{\|\vb*{P}^T\vb*{\tilde{q}^+}(x_1,x_2)\| \,\|\vb*{P}^T\vb*{\tilde{q}}(x_1,x_2) \|}{
    |\left<\vb*{\tilde{q}^+},\vb*{\tilde{q}}\right>|},
\end{align}
where the norm $\|\cdot\|$ measures the localized energy.
 Referred to \citet{luchini2014adjoint} for a comprehensive review.
In addition to quantifying the receptivity, the wavemaker also indicates the non-normality level of the flow field \citep{chomaz2005global}.
Initially introduced for base flows, \citet{meliga2016self} extended the application of wavemaker as a sensitivity analysis technique for mean flows.
We utilize this methodology to gauge the structural sensitivity of the RBF-FD-based global LNS operator, $\vb*{L}$.

\subsection{Mesh-free resolvent analysis (RA)}

The construction of the global Jacobian, $\vb*{L}$, leads to the direct discretization of the input-output system in equation (\ref{input_output}), yielding
\begin{align}\label{input_output_discrete}
    \vb*{\hat{u}}  = \vb*{H}(\omega)  \vb*{\hat{f}} ,
\end{align}
where
\begin{align}\label{discrete_resolvent}
   \vb*{H}(\omega)=\vb*{P}^T \vb*{C}\left(\mathrm{i}\omega \vb*{PP}^T-\vb*{L} \right)^{-1}\vb*{P} \vb*{B}
\end{align}
 is referred to as the discrete resolvent operator.

In the absence of nonlinear interactions described in table (\ref{summary_decomposition}), the application of input-output, or resolvent analysis (RA) provides a means to model them as optimal forcing inputs to the linear system in equation (\ref{input_output_discrete}).
The objective of resolvent analysis is to identify pairs of optimal forcings and their corresponding responses that maximize the gain, $\sigma^2$, defined as the ratio of the energy of the response to the energy of the forcing, 
\begin{align}\label{resolvent_energy}
    \sigma^2(\vb*{\hat{f}};\omega) = \frac{\|\hat{\vb*{u}}\|^2_{u}}{\|\vb*{\hat{f}}\|^2_f}=\frac{\left<\vb*{H}(\omega)\vb*{\hat{f}},\vb*{H}(\omega) \vb*{\hat{f}}\right>_u}{\left<\vb*{\hat{f}},\vb*{\hat{f}}\right>_f}.
\end{align} 
Refer to \citet{schmid2002stability} for a detailed discussion.
The energy of the response and the forcing are measured in the norms $\|\cdot\|_u$ and $\|\cdot\|_f$, induced by the inner products
\begin{align}\label{inner_prod}
    \left<\hat{\vb*{u}}_1,\hat{\vb*{u}}_2\right>_u =  \hat{\vb*{u}}_2^*\vb*{W}_u \hat{\vb*{u}}_1  \qquad \text{and} \qquad 
     \left<\vb*{\hat{f}}_1,\vb*{\hat{f}}_2\right>_f = \vb*{\hat{f}}_2^* \vb*{W}_f \vb*{\hat{f}}_1
\end{align}
on the output and input spaces, respectively.
Here, $\vb*{W}_u$ and $\vb*{W}_f$ are weight matrices containing both the numerical quadrature weights and weights associated with these inner products.

We further define the modified, or weighted, resolvent operator
\begin{align}
    \vb*{R}(\omega)\equiv\vb*{W}_u^{\frac{1}{2}} \vb*{H}(\omega)\vb*{W}_f^{-\frac{1}{2}}=\hat{\vb*{U}}\vb*{\Sigma}\hat{\vb*{F}}^*
\end{align}
to account for the energy in the input and output spaces. The optimal responses $\vb*{\hat{U}}=\mqty[{\hat{\vb*{u}}}_1,\cdots,{\hat{\vb*{u}}}_N]$ and the corresponding forcings $\vb*{\hat{F}}=\mqty[\vb*{\hat{f}}_1,\cdots,\vb*{\hat{f}}_N]$ are obtained from the singular value decomposition (SVD) of the modified resolvent operator and ranked by the energy gains $\Sigma = \mqty[\sigma_1,\cdots \sigma_N]$. The resulting modes are orthogonal in their respective inner products, that is, $\left<\hat{\vb*{u}}_j,\hat{\vb*{u}}_k\right>_u =  \left<\vb*{\hat{f}}_j,\vb*{\hat{f}}_k\right>_f =\delta_{j k}$. The optimal input and output modes are related through
\begin{align}\label{resolvent_discrete}
     \vb*{R}(\omega)\vb*{ \hat{f}}_j = \sigma_j(\omega) \hat{\vb*{u}}_j,
\end{align}
which provides a physical interpretation of the singular values and vectors. In practice, the optimal input forcings, $\vb*{\hat{f}}_j $, are determined as the solutions of the eigenvalue problem
\begin{align}
   \vb*{W}_f^{-1} \vb*{H}(\omega)^*\vb*{W}_u \vb*{H}(\omega)\vb*{\hat{f}}_j  =\sigma^2_j  \vb*{\hat{f}}_j,
\end{align}
where $\vb*{W}_f^{-1} \vb*{H}(\omega)^*\vb*{W}_u \vb*{H}(\omega)$ is Hermitian. The matrix inversion of $\left(\mathrm{i}\omega \vb*{P}\vb*{P}^T-\vb*{L}\right)$ is solved using LU-factorization \citep{sipp2013characterization}.

The RA-based wavemaker, 
\begin{align}\label{wavemaker_resolvent}
    \zeta_{\text{RA}}(x_1,x_2) = \frac{\|\vb*{\hat{u}}_1(x_1,x_2)\| \, \|\vb*{\hat{f}}_1(x_1,x_2)\| }{|\vb*{\hat{u}}_1^* \vb*{W}_u \vb*{\hat{f}}_1|},
\end{align}
is defined analogous to equation (\ref{structral_sens}). Within the framework of RA, it provides a quantitative measure of the effect of the localized feedback at any given frequency \citep{qadri2017effect,ribeiro2022wing, skene2022sparsifying,symon2018non}.

% \section{Accuracy and error analysis}

\section{Numerical implementations} \label{numerical}

\subsection{Node generation and parameter selection}\label{parameter_selection}

We employ the unstructured triangular mesh generator {\tt DistMesh} developed by \citet{persson2005mesh} to efficiently generate scattered nodes with localized refinement in regions of interest.
We highlight that the local connection information is not used in the computation.
The accuracy and stability of the PHS+poly RBF-FD method are contingent upon three parameters: the stencil size, $n$, the PHS exponent, $m$, and the polynomial degree, $q$.
we investigate the order of accuracy ranging from the standard choice of $q=2$ to the desired high-order accuracy for hydrodynamic stability analysis, here $q=4$.
Increasing the order of accuracy generally leads to a decrease in the size of the linear operator.

% In hydrodynamic stability analysis, it is desirable to have a compact stencil size to ensure numerical efficiency. We aim for a stencil size with $n\lesssim 40$, which leads to a constraint of the polynomial degree with $q\lesssim 4$.

% In hydrodynamic stability analysis, we focus on the order of accuracy ranging from the standard choice of $q=2$ to the desired high-order accuracy of $q=4$. The latter results in a reduction in the size of the linear operator.

% not going above this (compact stencil)

\begin{figure}[hbt!]
\centering
\includegraphics[trim = 0mm 0mm 0mm 0mm, clip, width=0.75\textwidth]{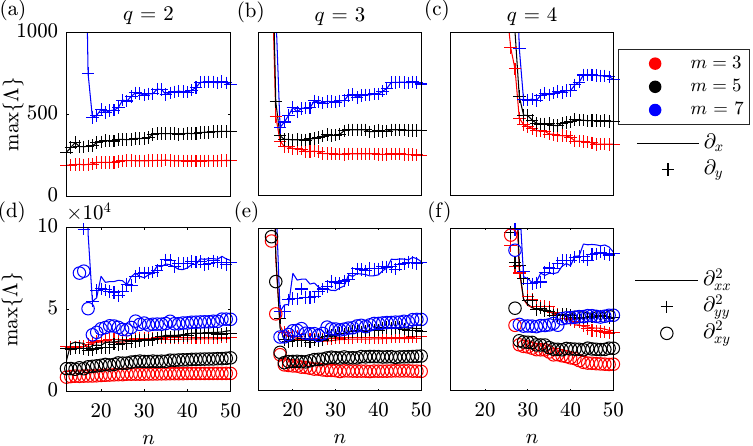}
\caption{Maximum local $\mathcal{L}$-Lebesgue function for different combinations of PHS exponents, $m$, and polynomial orders, $q$ for the first (a-c) and second derivatives (d-f) in the $x$- (solid line), $y$- ('+') and $xy$-directions (circle). }\label{Leb_function}
\end{figure}

The local $\mathcal{D}$-Lebesgue function proposed by \citet{shankar2018hyperviscosity} provides a measure of the eigenvalue stability of the RBF-FD method. For a given linear operator $\mathcal{D}$, the local Lebesgue function takes the form of the 1-norm of the RBF-FD weight in equation (\ref{Lf}), that is
\begin{align}
    \Lambda_\mathcal{D}(\vb*{x}_0; \{\vb*{x}_j\}_{j=1}^n) =  \|\vb*{w}\|_1 =\sum_{j = 1}^n |w_j|.
\end{align}
Larger values of the local Lebesgue function indicate an increased susceptibility to numerical instability in the assembled global differentiation matrix.
Figure \ref{Leb_function} shows the maximum value of the local Lebesgue function across various parameter combinations based on the test grid shown in figure \ref{mesh_cylinder}. 
Refer to the accompanying context of figure \ref{mesh_cylinder} for more details of the grid.
For all considered stencil sizes, polynomial degrees, and relevant operators, the minimax value is consistently achieved at $m=3$, suggesting its suitability for hydrodynamic stability analysis.

For two-dimensional problems, it has been suggested that the stencil size should satisfy $n\gtrsim(q+1)(q+2)$ \citep{flyer2016enhancing,flyer2016role}.
In practice, including a few additional nodes beyond the minimum requirement is often beneficial to improve performance. The specific choice of additional nodes may depend on the type of node sets used.
For instance, a recommended formula for Halton nodes is $n = (q+1)(q+2) + \lfloor{\ln{\left[(q+1)(q+2)\right]}}\rfloor$ \citep{le2023guidelines, shankar2018hyperviscosity}.
We here select the stencil size $n$ based on the assumption of a perfectly arranged hexagonal node distribution comprising $q/2+1$ layers, despite the actual distribution of nodes being heterogeneous. For practical examples, refer to figure \ref{nodes_local} and the surrounding context. The recommended parameter combinations are summarized in table \ref{parameters}.
In particular, we use $(n,m,q)=(37,3,4)$ for interior nodes and 
$(n,m,q)=(19,3,2)$ for nodes near boundaries in the construction of the discrete LNS operator, $\vb*{L}$. The RBF stencil for each node is determined as the nearest $n$ nodes acquired through a k-nearest neighbor (kNN) search.

% Refer to figure \ref{nodes_local} for illustrations.

\begin{table}[!htb]
      \centering
        \begin{tabular}{|c|c|c|c|c|}
\cline{2-5}
 \multicolumn{1}{c|}{}  & \citet{shankar2018hyperviscosity} & \citet{le2023guidelines} & present \\\hline\hline
 $q=2$ & $(n,m) = (14,3)$ &  $(n,m) = (14,5)$ &  $(n,m) = (19,3)$  \\ \hline
 $q=3$ & $(n,m) = (22,3)$ &  $(n,m) = (22,7)$ &  $(n,m) = (28,3)$ \\ \hline
 $q=4$ & $(n,m) = (33,3)$ &  $(n,m) = (33,7)$ &  $(n,m) = (37,3)$ \\ \hline
 % $q=5$ & $(n,m) = (45,5)$ &  $(n,m) = (45,9)$ &  -\\ \hline
        \end{tabular}
        \caption{Summary of parameter selections.}\label{parameters}
    \label{parameter}
\end{table}

% The practical implementation of the PHS+poly RBF-FDs follows \citet{flyer2016enhancing}.

\begin{figure}[hbt!]
\centering
\includegraphics[trim = 0mm 0mm 0mm 0mm, clip, width=.75\textwidth]{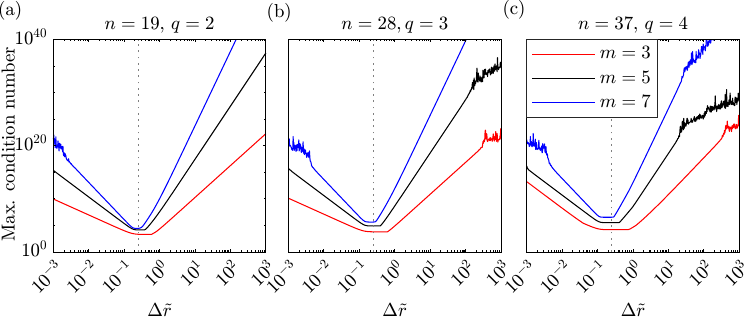}
\caption{Maximum condition number of $\vb*{A}_{\text{aug}}$ as a function of local grid spacing $\Delta \tilde{r}$: (a) $n=19,\, q=2$; (b) $n=28,\, q=3$; (c) $n=37,\, q=4$. The recommended local spacing of $\Delta \tilde{r} = 0.25$ is highlighted as dashed lines.}\label{cond_number}
\end{figure}

% \citet{le2023guidelines} examined the condition number for different parameter combinations, and \citet{shahane2021high} demonstrated that scaling the domain can improve the condition number.

The condition number of the augmented matrix $\vb*{A}_{\text{aug}}$, defined in equation (\ref{rbf_poly_compact}), affects the numerical stability in a similar way and hence needs to be considered separately.
The condition number for various parameter combinations has been investigated and discussed in \citet{le2023guidelines}.
 Independent of the parameter selection, \citet{shahane2021high} demonstrated that scaling the local stencil to a unit length scale can improve the condition number.
We further reveal the existence of an optimal range for the averaged local grid spacing, wherein the condition number of the augmented matrix $\vb*{A}_{\text{aug}}$ reaches its minimum value.

For a given stencil $\{{\vb*{x}}_j\}_{j=1}^n$, we perform scaling with respect to the location of $\vb*{x}_0$ with a scaling factor $\kappa$ such that
\begin{align}
  \{\tilde{\vb*{x}}_j\}_{j=1}^n  =  \{ \kappa ({\vb*{x}}_j -\vb*{x}_0 )\}_{j=1}^n,
\end{align}
and the averaged local grid spacing becomes $ \Delta \tilde{r} =\kappa \Delta r$.
Figure \ref{cond_number} shows the maximum condition number within the test grid as a function of $\Delta \tilde{r}$ for the considered parameters. 
Optimal performance is consistently achieved within the range of $0.15\lesssim\Delta \tilde{r} \lesssim 0.3$ across all examined parameter combinations.
Based on this, we perform spatial scaling to ensure an averaged local spacing of $\Delta \tilde{r} = 0.25$ before solving the RBF-FD weights in equation (\ref{Lf}).
The chain rule is applied to transform the RBF weights back to the original grid.

% Additionally, we reveal the existence of an optimal range for the averaged local grid spacing, $\Delta \tilde{r}$, wherein the condition number of the augmented matrix $\vb*{A}_{\text{aug}}$ reaches its minimum value, as shown in figure \ref{cond_number}.

\subsection{Boundary condition treatments}

\subsubsection{Homogeneous Dirichlet boundary conditions}

Homogeneous Dirichlet boundary conditions, $\vb*{q}'=0$, are widely employed in hydrodynamic stability analysis based on the assumption that perturbation variables either vanish at the solid wall or in the far field.
To approximate spatial derivatives for a node near such a boundary, we follow the Kansa method \citep{kansa1990multiquadrics2} and compute the RBF-FD weights using the local stencil consisting of both the interior nodes, $\{\vb*{x}^{(i)}\}_{j=1}^{n^{(i)}}$, and the boundary nodes, $\{\vb*{x}^{(b)}\}_{j=1}^{n^{(b)}}$.
Letting $g(\vb*{x}_j^{(b)})=0$, equation (\ref{Lf}) becomes
\begin{align}
   \mathcal{D} g(\vb*{x}_0)&= \sum_{j=1}^{n^{(i)}} w_{j}^{(i)} g(\vb*{x}_j^{(i)})+\sum_{j=1}^{n^{(b)}} w_{j}^{(b)} g(\vb*{x}_j^{(b)}) = \sum_{j=1}^{n^{(i)}} w_{j}^{(i)} g(\vb*{x}_j^{(i)}),
\end{align}
and only the weights for the interior nodes will be used in the computation.

\subsubsection{Boundary-normal derivatives}

% Neumann and other types of boundary conditions, such as inflow/outflow conditions, require boundary-normal derivatives and, therefore, special treatment for scattered nodes that are not structured near boundaries.

% Wall-normal derivatives commonly arise in hydrodynamic stability analysis due to various boundary conditions, such as Neumann conditions, inflow/outflow conditions, or the enforcement of continuity at the wall. 

% However, when dealing with scattered nodes that are not structured near the boundary, it becomes beneficial to determine a suitable local stencil for approximating these derivatives. 

Neumann conditions, inflow/outflow conditions, or the enforcement of continuity necessitate the evaluation of boundary-normal derivatives and require special treatment for scattered nodes that are not structured near boundaries.
To accurately approximate derivatives in the boundary-normal direction, we propose an elliptical stencil with its major axis perpendicular to the boundary.
The elliptical form is achieved by utilizing a weighted Euclidean distance metric 
\begin{align}
\tilde{r} = \sqrt{a^2+b^2}\sqrt{\frac{\left((\vb*{x}-\vb*{x}_0)\cdot \vb*{n}\right)^2}{a^2}+ \frac{\left((\vb*{x}-\vb*{x}_0)\cdot \vb*{t}\right)^2}{b^2}},
\end{align}
where $\vb*{n}$ and $\vb*{t}$ are the normal and tangential directions, respectively, and $a$ and $b$ denote the corresponding scaling factors. 
An eccentricity of $\frac{\sqrt{a^2-b^2}}{a}={\frac{2\sqrt{6}}{5}}$ is used in the computation.
We use the $q=2$ polynomial augmentation for the boundary-tangential direction and a higher-order of accuracy of $q=4$ for the boundary-normal direction.
An illustrative example is shown in figure \ref{nodes_local}(a).
This new strategy circumvents the need to increase the local stencil near domain boundaries to prevent the Runge phenomenon, as demonstrated in \citet{bayona2017role, bayona2019insight}. Without this special treatment, stable solutions for hydrodynamic stability analysis cannot be attained.

% By utilizing this specialized stencil, accurate and efficient approximation of derivatives in the wall-normal direction can be achieved.

% \begin{figure}[hbt!]
% \centering
% \includegraphics[trim = 10mm 10mm 10mm 10mm, clip, width=0.4\textwidth]{nodes_symmetric.pdf}
% \caption{

% Local stencil (red circle) for a given node (blue star) near the symmetric centerline (dot-dashed) that includes the interior nodes ($\{\vb*{x}^{(i)}\}_{j=1}^{n^{(i)}}$, black dot) and image ghost nodes ($\{\vb*{x}^{(i)}\}_{j=1}^{n^{}}$, green '+'). The counterparts of the image nodes, $\{\vb*{x}^{(s,s)}\}_{j=1}^{n^{}}$, are shown as yellow triangles. The area of each circle represents the corresponding local radial control volume, $\dd V_i$.
% }\label{mesh_symmetric}
% \end{figure}

\subsubsection{Symmetric and anti-symmetric boundary conditions}

\begin{figure}
\begin{minipage}{.5\linewidth}
\centering
{\includegraphics[trim = 10mm 10mm 10mm 0mm, clip, width=.85\textwidth]{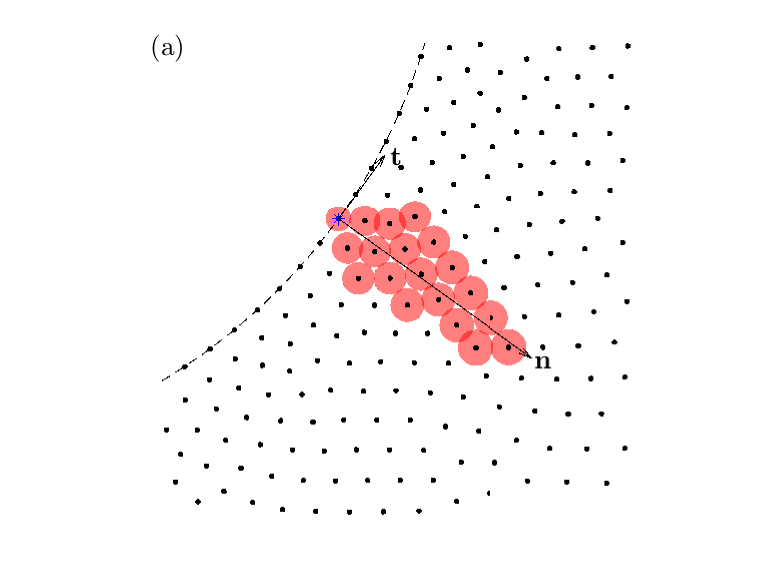}}
\end{minipage}%
\begin{minipage}{.5\linewidth}
\centering
{\includegraphics[trim = 10mm 10mm 10mm 0mm, clip, width=.85\textwidth]{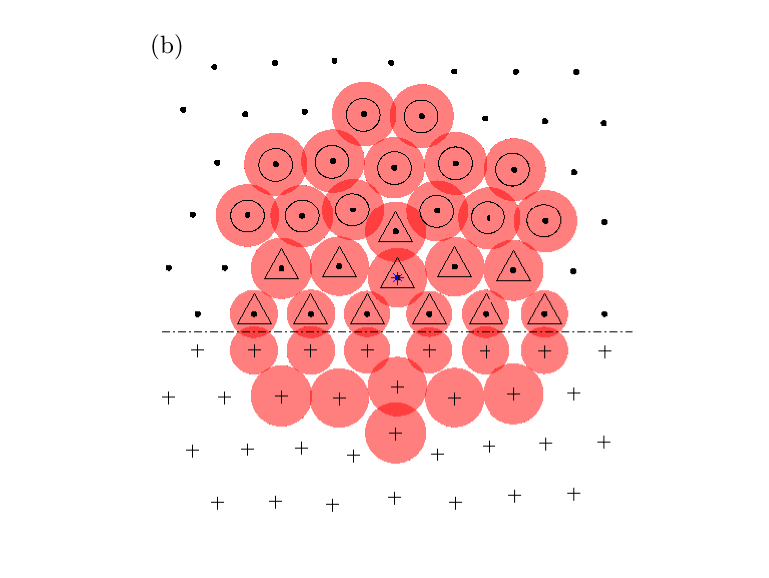}}
\end{minipage}
\caption{ 
RBF stencils (red-shaded circles) for a given node (blue star): (a) boundary-normal derivatives; (b) interior node near the symmetric centerline (dot-dashed). 
The latter stencil includes the interior nodes ($\{\vb*{x}^{\bullet}\}_{j=1}^{n^{\bullet}}$,  dot) and image ghost nodes ($\{\vb*{x}^{+}\}_{j=1}^{n^{+}}$, '+'). 
The counterparts of the image nodes, $\{\vb*{x}^{\Delta}\}_{j=1}^{n^{+}}$, and the remaining nodes, $\{\vb*{x}_j^{\circ}\}_{j=1}^{n^{\bullet}-n^{+}}$, are shown as triangles and circles, respectively. The area of each red-shaded circle represents the corresponding local radial control volume, $\dd V_i$.}
\label{nodes_local}
\end{figure}

To address pole singularities that arise in the polar coordinates and symmetric or anti-symmetric boundary conditions, we propose a generalization of the pole treatment method by \citet{mohseni2000numerical} for scattered nodes.
We introduce a set of ghost nodes, denoted as $\mathcal{X}^{+}$, which are symmetric to the interior nodes, $\mathcal{X}^{\bullet}$, with respect to the centerline.
For each interior node, we divide its local stencil into two disjoint sets such that 
$\{\vb*{x}\}_{j=1}^n = \{\vb*{x}_j^{\bullet}\}_{j=1}^{n^{\bullet}} \cup \{\vb*{x}_j^{+}\}_{j=1}^{n^{+}}$, where $\vb*{x}^{\bullet}\in \mathcal{X}^{\bullet}$ and $\vb*{x}^{+}\in \mathcal{X}^{+}$, respectively, and $n = n^{\bullet}+n^{+}$. Additionally, we define $\{\vb*{x}_j^{\Delta}\}_{j=1}^{n^{+}}\in \{\vb*{x}_j^{\bullet}\}_{j=1}^{n^{\bullet}}$ as the counterparts of the image nodes, and $\{\vb*{x}_j^{\circ}\}_{j=1}^{n^{\bullet}-n^{+}}=
\{\vb*{x}_j^{\bullet}\}_{j=1}^{n^{\bullet}} \setminus \{\vb*{x}_j^{\Delta}\}_{j=1}^{n^{+}}$ represents the remaining nodes within the interior nodes, 
Refer to figure \ref{nodes_local}(b) for detailed symbol explanations.
The function values at the image nodes are determined by their corresponding counterparts, given by 
\begin{align}
    g(\vb*{x}_j^{+})= \eta g(\vb*{x}_j^{\Delta}),
\end{align}
where $\eta$ depends on the type of boundary conditions being imposed. Specifically, we use $\eta=1$ for symmetric boundary conditions and $\eta=-1$ for anti-symmetric boundary conditions. The RBF-FD weights in equation (\ref{Lf}) can be written as
\begin{align}
   \mathcal{D} g(\vb*{x}_0)&= \sum_{j=1}^{n^{\bullet}} w_{j}^{\bullet} g(\vb*{x}_j^{\bullet})+\sum_{j=1}^{n^{+}} w_{j}^{+} g(\vb*{x}_j^{+}) = \sum_{j=1}^{n^{\bullet}-n^{+}} w_{j}^{\circ} g(\vb*{x}_j^{\circ})+\sum_{j=1}^{n^{+}} \left(\eta w_{j}^{+}+w_{j}^{\Delta}\right) g(\vb*{x}_j^{\Delta}).
\end{align}
This treatment allows us to approximate derivatives solely based on the function values at the interior nodes, effectively handling pole singularities and addressing the challenges associated with scattered nodes.

% \newpage \clearpage

\section{Applications} \label{applications}

\begin{table}[!htb]
      \centering
    \renewcommand{\arraystretch}{1.2}
        \begin{tabular}{|c|c|c|c|c|c|c| }
      \cline{1-7}
    Flow   & $\mathrm{Re}$ & $M$ & Flow type & Base state & Analysis & Sec.\\
 \hline 
 \hline
 Cylinder wake \citep{chu2023rbf}  & $47- 180$ & - & 2D laminar   & $\overline{u},\overline{v}$ & LST/RA/WM &  \S \ref{cylinder wake}\\ 
 \hline
  Boundary layer (ZPG)  & $6\times 10^5$ & - & 2D laminar  & $u_{\text{ZPG}}, v_{\text{ZPG}}$ & RA/WM & \S \ref{boundary layer}\\ 
 \hline
   Jet \citep{bresetal_2018jfm} & $\approx 10^6$ & 0.9 & 3D turbulent & $\overline{\rho},\overline{u}_x,\overline{u}_r,\overline{u}_{\theta} ,\overline{T}$ & RA/WM  &\S \ref{jet}\\ 
 \hline
        \end{tabular}
        \caption{Overview of datasets and analyses. The columns from left to right indicate the flow description, Reynolds number, Mach number, flow type, base state, analysis type, and section number.
        The zero-pressure-gradient (ZPG) Blasius solution is used for analyzing the boundary layer. Analyses include linear stability (LST) and resolvent analyses (RA), along with wavemaker (WM).}
    \label{flow_summary}
\end{table}

 We demonstrate the mesh-free RBF-FD-based hydrodynamic stability framework outlined in \S \ref{meshfree_stability} using three representative examples: canonical steady and unsteady cylinder wakes, a self-similar non-parallel steady laminar boundary-layer flow, and the turbulent mean of a transonic jet, as summarized in table \ref{flow_summary}. 
 These three examples are benchmark problems for open flows and are appropriate for 
 validating mesh-free hydrodynamic stability.

 % We highlight that these are carried out solely for the purpose of demonstration and validation of the proposed mesh-free approach.

% These open flows

\subsection{Cylinder wake}\label{cylinder wake}

We first consider the incompressible cylinder flow at diameter-based Reynolds numbers, $\mathrm{Re}= \frac{U_{\infty} D}{\nu}$, ranging from 47 to 180 and investigate the mean-flow stability within the two-dimensional laminar regime. The occurrence of the periodic von K{\'a}rm{\'a}n vortex shedding in the cylinder wake beyond the critical Reynolds number of $\mathrm{Re_c}\simeq 47$ is a well-known phenomenon, owing to a Hopf bifurcation that results in flow instability, see, e.g., \citep{benard1908formation,von1911mechanismus}. Beyond the limit of $\mathrm{Re} \simeq 188$, the cylinder flow becomes three-dimensional \citep{barkley1996three, williamson1996vortex}.

Classical LST analysis of the cylinder base flow ought to predict the onset of unsteadiness \citep{jackson1987finite,zebib1987stability} 
but fails to capture the vortex-shedding frequency beyond $\mathrm{Re}_c$ \citep{barkley2006linear,sipp2007global}.
Previous studies by \citet{hammond1997global} and \citet {pier2002frequency} show that LST analysis around the cylinder mean flow accurately identifies the vortex-shedding frequency compared to experimental measurements.
\citet{barkley2006linear} supported these findings and showed that the cylinder mean flow is marginally stable in the 2D regime.
\citet{sipp2007global} subsequently provided theoretical underpinning by conducting a weakly nonlinear analysis and establishing criteria for utilizing mean flows in LST analysis.
The vortex-shedding dynamics beyond the critical Reynolds number were later investigated using a self-consistent model \citep{mantivc2014self,mantivc2015self} and RA \citep{jin2021energy,symon2018non} based on the mean flow stability. We here conduct both LST and resolvent analyses of the cylinder wake mean flow.

% and the unsteady flow structures \citep{barkley2006linear, pier2002frequency}.

% \citet{symon2018non} further demonstrated that the leading linear stability and resolvent modes for the cylinder wake are nearly identical. 

 % code, we choose the now classical mean-flow stability analyses of the cylinder wake by \citet{pier2002frequency} and \citet{barkley2006linear}.

\begin{figure}[hbt!]
\centering
\includegraphics[trim = 5mm 5mm 10mm 10mm, clip, width=0.45\textwidth]{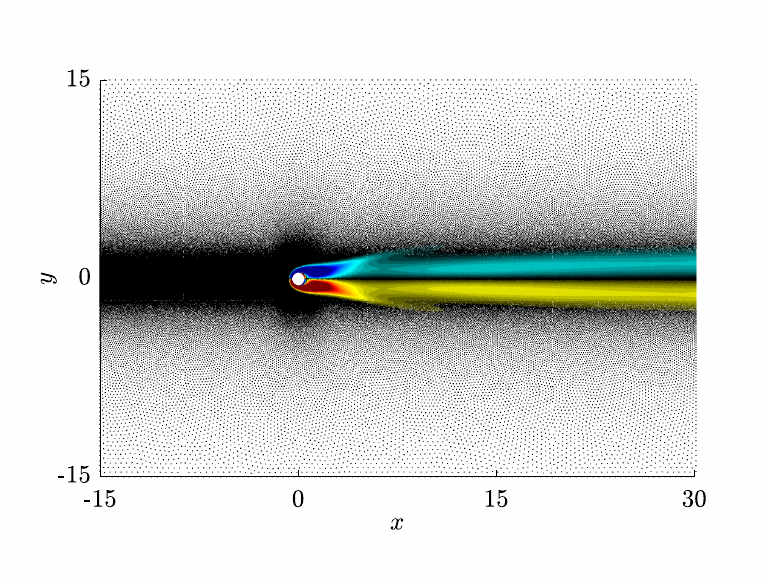}
\caption{Computational grid for cylinder flows with $N=118225$ nodes, colored using the mean vorticity, $\overline{\vb*{\omega}}=\vb*{D}_x \overline{\vb*{v}}-\vb*{D}_y \overline{\vb*{u}}$, at $\mathrm{Re}=100$. }\label{mesh_cylinder}
\end{figure}

We define the the computational domain $\Omega$ as the exterior of the cylinder $r\geq D/2 =0.5$ and within the rectangle $-15\leq x\leq  30,\, -15\leq y\leq 15$.
The computational domain is discretized using $N=118225$ scattered nodes.
Local grid refinement is employed near the cylinder with a characteristic distance of $\Delta r=0.03$ and around the wake centerline with $\Delta r=0.04$ to better resolve the flow structures.
The unsteady cylinder flow is simulated using the PHS+poly RBF-FD version of the fractional-step, staggered-grid incompressible Navier-Stokes solver by \citet{chu2023rbf}.
The mean-flow profiles are obtained as the time average of flow over 20 vortex-shedding cycles. 
Figure \ref{mesh_cylinder} shows the mean vorticity at $\mathrm{Re}=100$ and the computational grid.
Homogeneous boundary conditions, $u'= v' =0$, are prescribed at the inlet and the cylinder surface.
Symmetric boundary conditions with $v'=\partial u'/\partial y =0$ are applied at the transverse boundaries.
A stress-free outflow condition, $-p'\vb*{n}+\frac{1}{\mathrm{Re}}{\nabla}\vb*{u}'\cdot\vb*{n}=0$, where $\vb*{n}=[1, 0]^T$ is the outflow direction, is enforced at the outflow.

The local wavemaker sensitivity in equation (\ref{structral_sens}) and the resolvent gain in equation (\ref{resolvent_energy}) and 
are both quantified in terms of the perturbed kinetic energy.
To this end, we define the integration matrices
\begin{align}\label{W_kinetic}
    \vb*{W}_u=\vb*{W}_f\equiv \mqty[ \vb*{1} & \vb*{0}  \\  \vb*{0} & \vb*{1} ] \otimes \mathrm{diag}(\dd V_1, \dd V_2,\cdots, \dd V_N)
\end{align}
in equation (\ref{inner_prod}) to approximate the integral within the computational domain, where $\otimes $ denotes the Kronecker product. 
Here, $\dd V_i = \xi \pi(\Delta r_i) ^2$ is the local radial control volume for each grid. The constant $\xi$ ensures consistency with the total control volume and is defined by letting
$\sum_{i=1}^N\dd V_i = \xi \pi \sum_{i=1}^N (\Delta r_i) ^2 = \int_{\Omega}1\dd \vb*{x}$.
Refer to figure \ref{nodes_local} for practical illustrations.

% \newpage\clearpage

% \subsubsection{Mean-flow stability analysis} \label{Re_47}

\begin{figure}[hbt!]
\centering
\includegraphics[trim = 0mm 0mm 5mm 0mm, clip, width=0.5\textwidth]{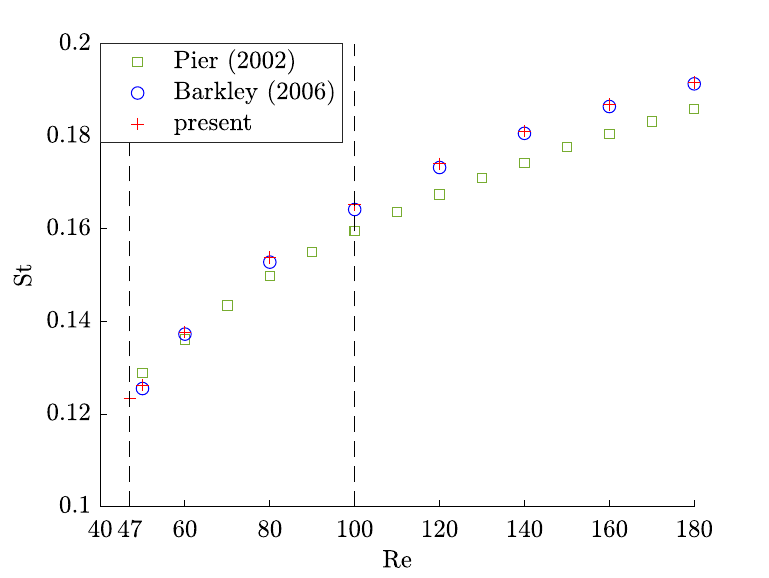}
\caption{Vortex shedding frequency predicted by the leading eigenvalue of the mean-flow stability problem as a function of Reynolds number. The frequency is given as the non-dimensional Strouhal number $\mathrm{St}=\lambda_i/2\pi$. Shown for comparison are results from \citet{pier2002frequency} (green square) and \citet{barkley2006linear} (blue circle). Two representative Reynolds numbers for the following analysis, $\mathrm{Re}_c \approx 47$ and $\mathrm{Re}=100$, are highlighted as dashed lines.
}\label{eigen_f}
\end{figure}

Figure \ref{eigen_f} shows the Strouhal number, $\mathrm{St}=\lambda_i/2\pi$, associated with the leading eigenvalues at varying Reynolds numbers.
Starting from the critical Reynolds number of $\mathrm{Re}_c \approx 47$, the frequency-Reynolds number dependence exhibits the typical features of a Hopf bifurcation.
Our results are in good agreement with \citet{barkley2006linear}, and similarly, 
deviate no more than $3\%$ from those reported by \citet{pier2002frequency}. 
The maximum growth rates are almost identical to zero (see also figure \ref{spectrum_cylinder} below), confirming that the mean flow is marginally stable.

% \newpage\clearpage

% \subsubsection{Comparative study of linear stability and resolvent analysis at $\mathrm{Re}_c \approx 47$ and $\mathrm{Re}=100$} \label{Re_100}

\begin{figure}[hbt!]
\centering
\includegraphics[trim = 0mm 20mm 0mm 22mm, clip, width=.8\textwidth]{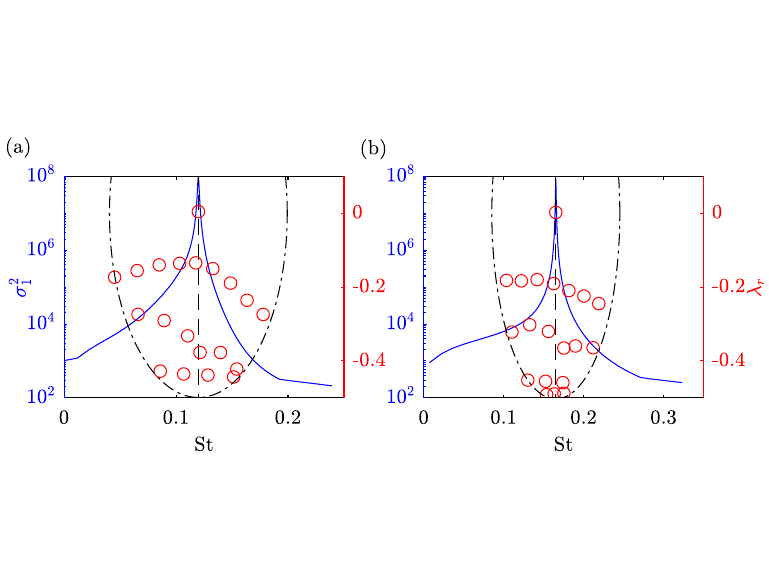}
\caption{Resolvent singular values (blue curve) and stability eigenvalue (red circle) spectra for $\mathrm{Re} = 47$ (a) and $\mathrm{Re} = 100$ (b). The 20 eigenvalues closest to 
$\lambda_c=0+0.753\mathrm{i}$ or $\mathrm{St}_c=0.1199$ and
$\lambda=0+1.038\mathrm{i}$ or $\mathrm{St}=0.1652$ (dashed lines)
 were found within the regions outlined by the black dot-dashed lines using the shift-and-invert Arnoldi algorithm for $\mathrm{Re} = 47$ and $100$, respectively.
}\label{spectrum_cylinder}
\end{figure}

% Before studying the modes in more detail, we first examine the resolvent singular value and stability eigenvalue spectra to confirm that peaks in the resolvent gain are associated with an eigenvalue. 

We next conduct a comparative study of LST and RA at two representative Reynolds numbers, the critical Reynolds number of $\mathrm{Re}=47$ and $\mathrm{Re}=100$, as an example of the unsteady regime.
Figure \ref{spectrum_cylinder} shows the resulting resolvent singular value and stability eigenvalue spectra.
At $\mathrm{Re}=\mathrm{Re}_c$, both the peak of the resolvent gain and the leading eigenvalue identify the same frequency, $\mathrm{St}_c=0.1199$, as the vortex shedding frequency. This value is in good agreement with the results of 
the LST analysis around the base flow, see, e.g.,
\citet{giannetti2007structural} ($\mathrm{St}_c\simeq0.118$ for $\mathrm{Re}_c\simeq 46.7$),
\citet{marquet2008sensitivity} ($\mathrm{St}_c\simeq0.116$ for $\mathrm{Re}_c\simeq 46.8$), and \citet{sipp2007global} ($\mathrm{St}_c\simeq0.118$ for $\mathrm{Re}_c\simeq 46.6$).
Similarly, the resolvent singular value spectrum at $\mathrm{Re} = 100$ in panel \ref{spectrum_cylinder}(b) displays a clear peak at the vortex shedding frequency, which is now at $\mathrm{St}=0.1652$, and again coinciding with the least stable LST eigenvalue. 
This result matches closely with the findings of earlier Strouhal-Reynolds number relationship \citep{fey1998new,jiang2017strouhal, mantivc2014self, ponta2004strouhal,williamson1996vortex}
and RA \citep{jin2021energy, symon2018non} studies.

% \newpage\clearpage

\begin{figure}[hbt!]
\centering
\includegraphics[trim = 0mm 5mm 0mm 7mm, clip, width=.8\textwidth]{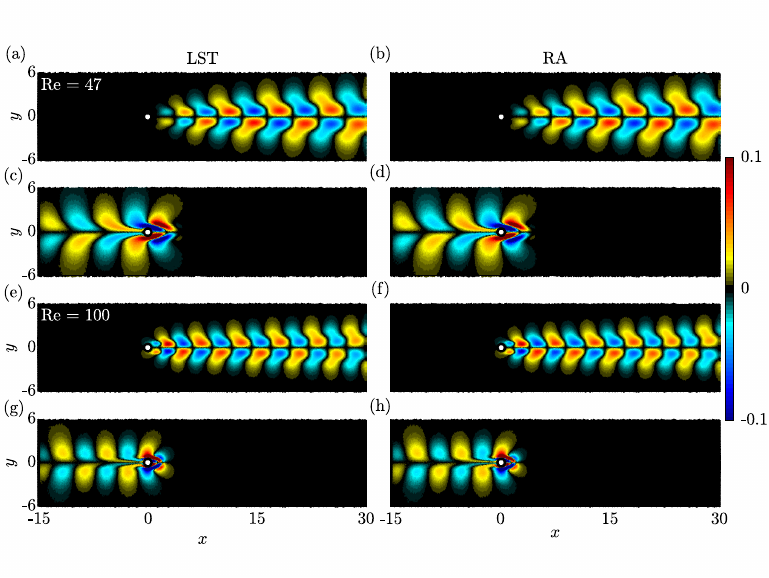}
\caption{Leading modes for $\mathrm{Re} = 47$ at $\mathrm{St}=0.1199$ (a-d) and $\mathrm{Re} = 100$ at $\mathrm{St}=0.1652$ (e-h): (a,e) LST modes; (b,f) response modes; (c,g) adjoint modes; (d,h) forcing modes.
}\label{modes_cylinder}
\end{figure}

Figure \ref{modes_cylinder} shows the leading LST and RA modes for the cylinder mean flow at $\mathrm{Re} = \mathrm{Re}_c$ and $\mathrm{Re} = 100$.
The leading LST and optimal response modes, along with their corresponding adjoint and forcing modes, are near-identical at both Reynolds numbers.
The resemblance observed between the LST and resolvent response modes is to be anticipated, as the singular value of the resolvent attains its peak at the vortex-shedding frequency \citep{beneddine2016conditions,symon2018non}, thus exhibiting the characteristic vortex-shedding structure.
This implies that the optimal forcing leverages the global instability mode to achieve maximum gain. 
Differing from the optimal response and LST modes, the optimal forcing and adjoint modes are active far upstream of the cylinder but peak downstream in close vicinity to the cylinder.
This hallmark of convective instability was similarly observed in previous works
\citep{jin2021energy,marquet2008sensitivity,symon2018non}. 

% The similarity between the stability and resolvent response modes is not unexpected as the singular value of the resolvent peaks at the vortex shedding frequency \citep{beneddine2016conditions,symon2018non} and exhibits the typical vortex-shedding structure. 

% \newpage \clearpage

% \subsubsection{structural sensitivity}

\begin{figure}[hbt!]
\centering
\includegraphics[trim = 0mm 16mm 0mm 11mm, clip, width=0.75\textwidth]{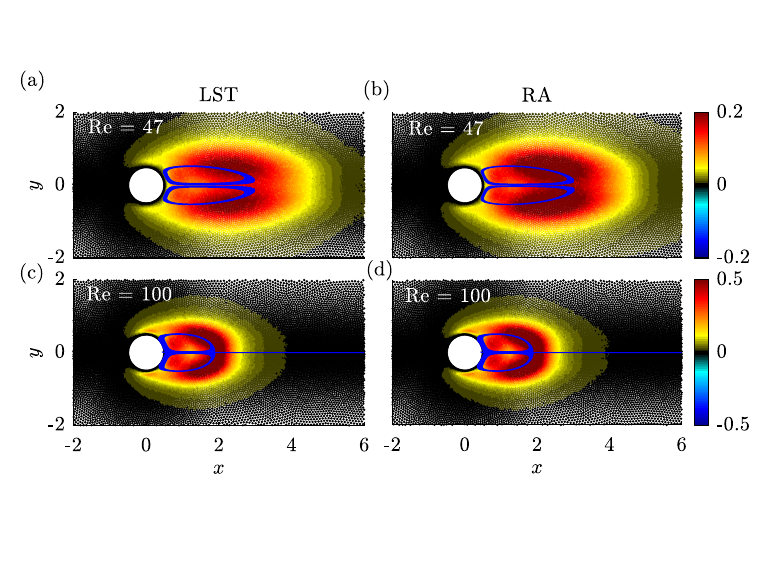}
\caption{ Wavemakers $\zeta_{\text{LST}}$ for mean flow at (a) $\mathrm{Re}=47$ and (c) $\mathrm{Re}=100$. Respective results for $\zeta_{\text{RA}}$ are shown in (b) and (d). The blue curve represents the mean flow streamlines.}
\label{wavemaker_cylinder}
\end{figure}

We finally investigate the wavemaker $\zeta_{\text{LST}}$, defined in equation (\ref{structral_sens}), in figure \ref{wavemaker_cylinder} to quantify the sensitivity of spatially localized feedback. 
Wavemakers obtained from LST and RA look similar for both Reynolds numbers and reach their maxima in two symmetrically positioned lobes located across the separation bubble.
This result signifies the promising applicability of RA-based wavemakers in accurately identifying the region where the flow instability mechanisms happen.
The wavemaker patterns at the critical Reynolds number shown in panels \ref{wavemaker_cylinder} (a,c) compare well to those previously reported by \citet{giannetti2007structural} and \citet{marquet2008sensitivity}.
The shrinking of wavemaker region at $\mathrm{Re}=100$ in panels \ref{wavemaker_cylinder} (b,d) closely matches with the findings in \citet{meliga2016self} and \citet{symon2018non}.

% \newpage \clearpage

\subsection{Blasius boundary layer}\label{boundary layer}

We now examine the incompressible two-dimensional non-normal flat-plate boundary layer, a classic example of convectively unstable flows characterized by the amplification of disturbances during downstream advection.
    The convective instability of boundary-layer flows has been extensively studied through 1D LST analysis over the past century, e.g., \citep{jordinson1970flat, mack1976numerical,mack1984boundary,reed1996linear, smith1979non}. Above the critical displacement-thickness Reynolds number of $\mathrm{Re}_{\delta,c} \approx 520$, the Tollmien–Schlichting (TS) waves are known to arise as unstable eigenmodes of the Orr-Sommerfeld equation \citep{butler1992three}.     
    While these locally unstable waves are damped in 2D LST \citep{aakervik2008global,alizard2007spatially, bagheri2009matrix, ehrenstein2005two}, investigations into the non-normality of the linearized Navier-Stokes equations for open flows have revealed an alternative pathway for disturbance amplification \citep{chomaz2005global,marquet2009direct,sipp2010dynamics}.
    To analyze the non-modal behavior of boundary-layer flows, input-output analysis has been conducted to determine the optimal harmonic forcing that results in the largest asymptotic response \citep{alizard2009sensitivity, bagheri2009inputb,bagheri2009input,monokrousos2010global,schmid2002stability,sipp2013characterization}.

Following the setup by \citet{sipp2013characterization}, we compute the resolvent gain in the restricted domain $\Omega_\text{RA}=[x,\,y/\delta] \in [0.02,\,1]\times[0,\,22.52]$ within the computational domain $\Omega=[x,\,y/\delta] \in [0.02,\,1.27]\times[0,\,22.52]$
such that the forcing optimizes the ratio between the restricted kinetic energy, $\iint_{\Omega_\text{RA}} \left(|\hat{u}|^2+|\hat{v}|^2\right)  \,\dd x \dd y$, and the integral of forcing, $\iint_{\Omega} \left(|\hat{f}_u|^2+|\hat{f}_v|^2\right)  \,\dd x \dd y$. 
The zero-pressure-gradient (ZPG) asymptotic Blasius solution, characterized by a local Reynolds number of $\mathrm{Re}_x = \frac{U_{\infty}x}{\nu}= 6\times 10^5$ or $\mathrm{Re}_{\delta}= \frac{U_{\infty}\delta(x)}{\nu} =1332$ at the end of the restricted domain $\Omega_\text{RA}$, is used as the base flow to investigate the flow instabilities, where $\delta(x) = 1.72\sqrt{x/\mathrm{Re}_x}$ is the displacement thickness.
The leading edge ($x<0.02$) is removed to avoid the singularity of the self-similar solution.
Hereafter, we use the notations $\mathrm{Re}=\mathrm{Re}_{x=1}$ and $\delta = \delta(x=1)$ for simplicity.

% The flat-plate boundary layer is characterized by a local Reynolds number of $\mathrm{Re}_x = \frac{U_{\infty}x}{\nu}= 6\times 10^5$ or $\mathrm{Re}_{\delta}= \frac{U_{\infty}\delta(x)}{\nu} =1332$ at the end of the restricted domain $\Omega_\text{RA}$.

% Following the setup by \citet{sipp2013characterization}, we here conduct the mesh-free RA for a flat-plate boundary layer characterized by a local Reynolds number of $\mathrm{Re}_x = \frac{U_{\infty}x}{\nu}\in \left[0, 6\times 10^5 \right]$ or $\mathrm{Re}_{\delta}= \frac{U_{\infty}\delta(x)}{\nu} \in \left[0, 1332\right]$ for $x\in\left[0,1\right]$.

% \begin{figure}[hbt!]
% \centering
% \includegraphics[trim = 0mm 5mm 0mm 5mm, clip, width=0.8\textwidth]{modes_Blasius_local.pdf}
% \caption{Local computational grid, colored using the optimal response (a) and forcing(b). }\label{modes_Blasius_local}
% \end{figure}

The computational domain $\Omega$ is discretized using $N=726143$ scattered nodes, corresponding to $2.2\times10^6$
degrees of freedom.
In comparison,  \citet{sipp2013characterization} employ $13.7\times10^6$ degrees of freedom. The characteristic distances of the grid are $\Delta r=0.069\delta$ near the flat plate ($y/\delta <4$), $\Delta r=0.077\delta$ near the inlet ($x < 0.025$), and average at $0.081\delta$ over the whole domain. 
At the inlet and the flat plate, we prescribe $u'=v'=0$. A symmetric boundary condition with $v'=\partial u'/\partial y =0$ is applied at the far-field ($y/\delta =22.52$). A stress-free outflow condition is enforced at the outflow.
 The weight matrix $\vb*{W}_u$ is defined in equation (\ref{W_kinetic}), but with zero weights in the region $\Omega \setminus \Omega_\text{RA}$.

% \newpage \clearpage

 \begin{figure}[hbt!]
\centering
\includegraphics[trim = 0mm 0mm 0mm 0mm, clip, width=0.5\textwidth]{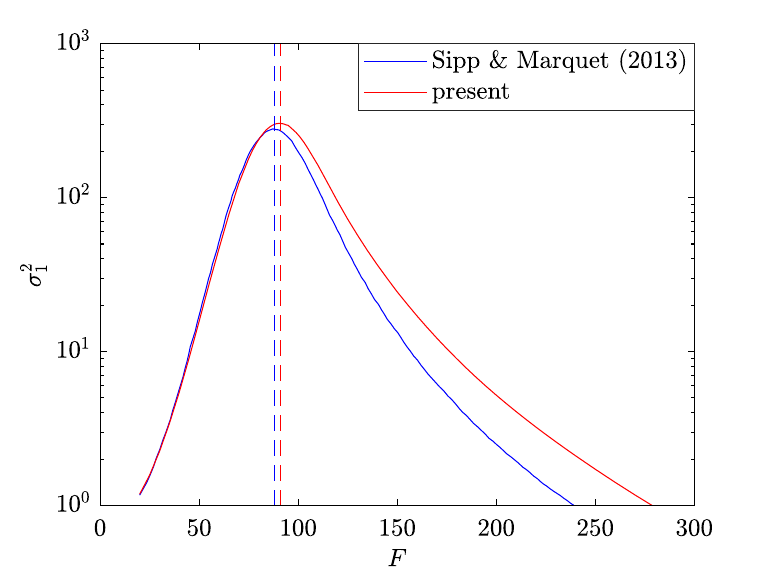}
\caption{
Resolvent gains (solid lines) and peak frequencies (dashed lines) for the flat-plate boundary layer and $\mathrm{Re} = 6\cdot 10^5$.}\label{resolvent_spectra_BL}
\end{figure}

We first examine the leading resolvent singular value as a function of the normalized frequency, $F = 10^6\cdot \omega/ \mathrm{Re}$, in figure \ref{resolvent_spectra_BL}.
Overall, our results agree well with those reported by \citet{sipp2013characterization}.
 The slight deviation of the peak and the resolvent gains beyond the peak ($F\gtrsim 88$) are most likely attributed to differences in the base flows, that is, the ZPG self-similar Blasius solution in the present study and the fully non-parallel numerical data of \citet{sipp2013characterization} that includes the leading edge.
 The discrepancy at higher frequencies is potentially also related to the truncation of the leading edge, where the asymptotic Blasius solution becomes singular.

% \newpage \clearpage

\begin{figure}[hbt!]
\centering
\includegraphics[trim = 5mm 0mm 0mm 0mm, clip, width=.75\textwidth]{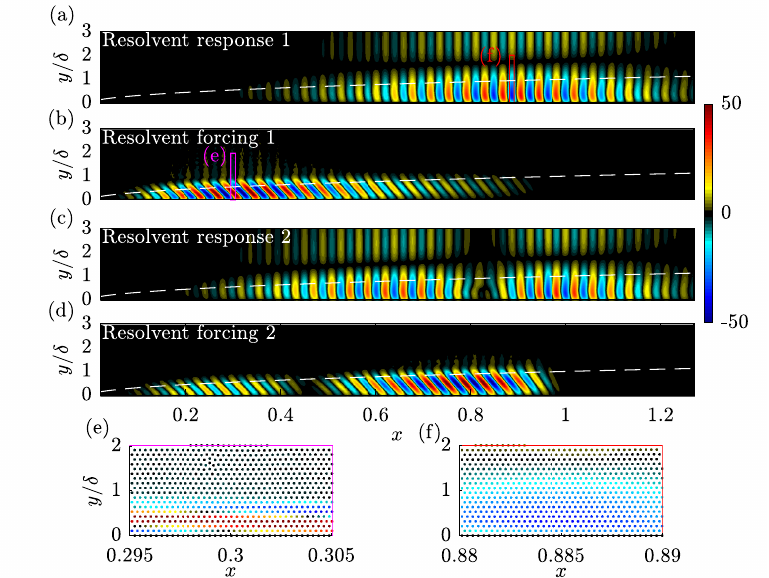}
\caption{Optimal and suboptimal resolvent forcings (b,d) and corresponding responses (a,c) for the flat-plate boundary layer at $F=10^6\cdot \omega/ \mathrm{Re}=100$. 
The normalized stream-wise velocity components have been interpolated onto a stretched Cartesian mesh for visualization. 
Panels (e) and (f) show the local regions of the optimal forcing (magenta box) and corresponding response modes (red box) with the largest magnitudes, respectively, on the scattered nodes used for the 
computation.
The Blasius displacement thickness, $\delta(x)$, is highlighted as the white dashed line.
}\label{modes_Blasius}
\end{figure}

Figure \ref{modes_Blasius} shows the optimal and suboptimal forcings and corresponding responses at $F = 100$.
The optimal response exhibits clear Tollmien–Schlichting (TS) wavepackets in the downstream region, while the upstream tilted structures in the optimal forcing highlight the active role of the Orr mechanism in extracting energy from the mean shear via the Reynolds stress \citep{butler1992three}.
The clear spatial separation between leading resolvent forcing and response modes indicates the stream-wise non-normality of the system \citep{chomaz2005global,marquet2008sensitivity,sipp2010dynamics}.
The suboptimal forcing and response are similar to the leading modes except for two local maxima.
This modulation confirms the orthogonality in their respective inner products.
Qualitative comparisons with previous studies by \citet{aakervik2008global,monokrousos2010global,brandt2011effect} and \citet{sipp2013characterization} verify the capability of the present framework in identifying the convective instability of the boundary-layer flow.

% \newpage \clearpage

 \begin{figure}[hbt!]
\centering
\includegraphics[trim = 0mm 0mm 0mm 0mm, clip, width=0.6\textwidth]{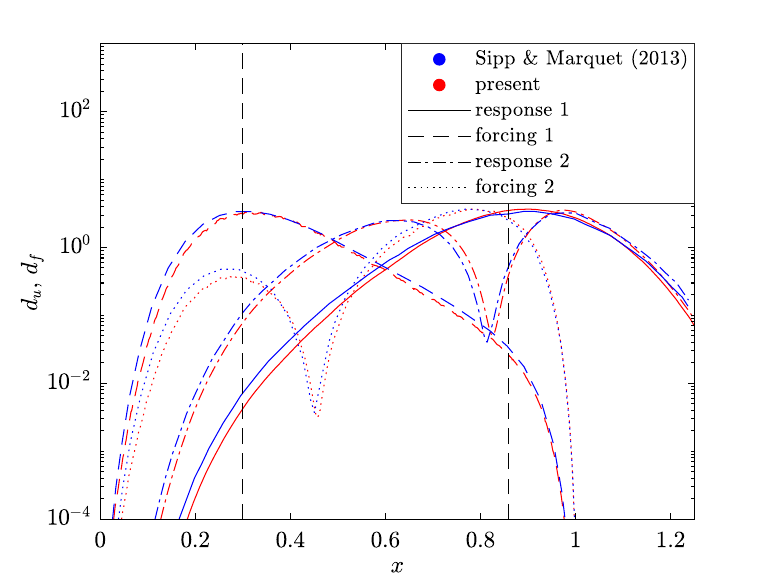}
\caption{Spatial distributions of energy density for optimal (dashed) and suboptimal (dotted) forcings and corresponding responses (solid and dot-dashed, respectively) at $F = 100$. 
The results (red) are compared to those reported by \citet{sipp2013characterization} (blue).
The two vertical solid lines represent the upstream neutral point (branch I) and the downstream neutral point (branch II) from a local stability analysis. }\label{energy_density}
\end{figure}

As a quantitative assessment of the flow structures, 
we examine the energy density functions,
 \begin{align}
     d_f(x)= \int_0^{y_\text{max}} \left(|\hat{f}_u|^2+|\hat{f}_v|^2\right)  \,\dd y ,\qquad \text{and} \qquad d_u(x)= \int_0^{y_\text{max}} \left(|\hat{u}|^2+|\hat{v}|^2\right)  \,\dd y,
 \end{align}
in figure \ref{energy_density} for the modes shown in figure \ref{modes_Blasius}.
Our results are almost identical to \citet{sipp2013characterization}
The spatial distribution of the optimal forcing unambiguously identifies the location of the upstream neutral point (branch I) from a local stability analysis at $x=0.3$ 
and the corresponding response is localized at $x=0.89$, which is in close proximity to the downstream neutral point (branch II).

% The accuracy of the present framework is validated by the almost identical results obtained for both optimal and suboptimal modes compared to those reported by \citet{sipp2013characterization}.

% \newpage \clearpage

 \begin{figure}[hbt!]
\centering
\includegraphics[trim = 0mm 18mm 0mm 15mm, clip, width=.85\textwidth]{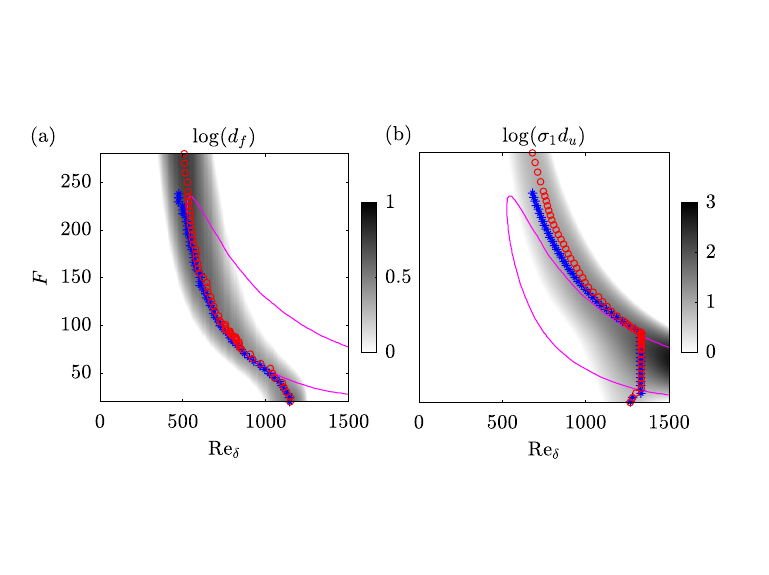}
\caption{Energy density distributions for the optimal forcings (a) and responses (b) as a function of frequency.
The locations of maximum energy densities are marked as red circles.
The results reported by \citet{sipp2013characterization} are shown as blue stars for comparison. 
The neutral curve obtained from a local LST analysis is shown as the magenta line.}\label{energy_density_frequency}
\end{figure}

Figure \ref{energy_density_frequency} shows the energy density distributions of the optimal forcing and response as a function of frequency, with the stream-wise coordinate given in terms of the local Reynolds number, $\mathrm{Re}_{\delta}$.
The optimal forcings agree well with \citet{sipp2013characterization}, and their maxima effectively delineate the convectively stable/unstable boundary (branch I) obtained using local stability theory. The Orr and TS mechanisms coexist and compete while both contribute to the overall energy gain of RA.
As frequency increases, the spatial support of the TS-like optimal responses decreases, suggesting that the TS mechanism is only supported in a limited region of the shear layer at high frequencies.
Additionally, the energy density of the forcing shows an increasing trend, indicating that the Orr mechanism becomes dominant at high frequencies.
Similar to the deviation in the resolvent gain spectra shown in figure \ref{resolvent_spectra_BL}, we observe a slight downstream shift in the peak of optimal responses compared to the results reported \citet{sipp2013characterization} for increasing frequency ($F\gtrsim 130$).

% Overall, good agreement with a slight shift towards ...

% However, these discrepancies are minor.
% The above qualitative comparisons validate the accuracy of the mesh-free RA approach.

% \newpage \clearpage

\begin{figure}[hbt!]
\centering
\includegraphics[trim = 0mm 0mm 0mm 5mm, clip, width=0.7\textwidth]{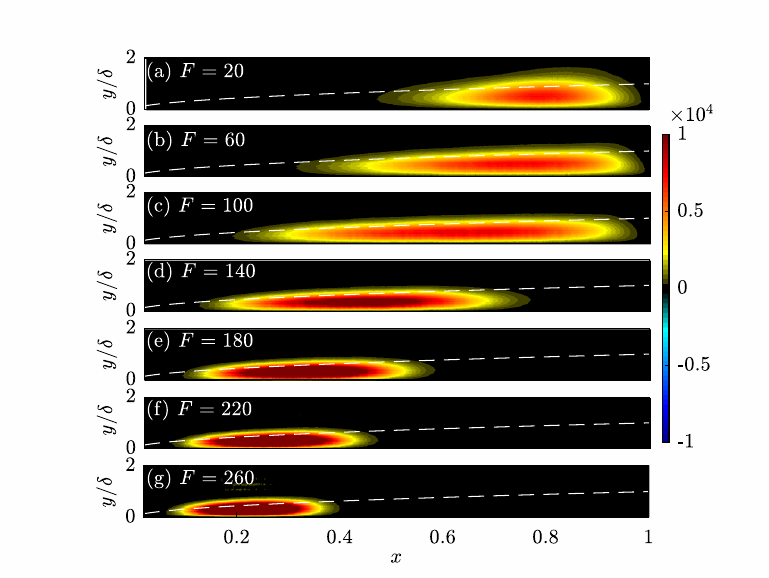}
\caption{RA-based wavemaker $\zeta_{\text{RA}}$ for the flat-plate boundary layer as a function of frequency.}
\label{wavemaker_BL}
\end{figure}

Finally, we show the RA-based wavemaker, defined in equation (\ref{wavemaker_resolvent}), in figure \ref{wavemaker_BL} to examine the instability mechanisms at different frequencies. 
The maximum magnitude of $\zeta_{\text{RA}}$ shows an increasing trend with frequency.
The large magnitudes of the wavemaker are attributed to the fact that the value of the term $|\vb*{\hat{u}}_1^* \vb*{W}_u \vb*{\hat{f}}_1|^{-1}$ is considerably greater than the value of the self-adjoint modes, which is equal to 1 \citep{marquet2009direct}.
This again confirms that the boundary-layer flow exhibits high non-normality.
For $F\lesssim 100$ (panels \ref{wavemaker_BL} (a-c)), the RA-based wavemaker exhibits an elongated shape and is artificially restricted within the optimization domain $\Omega_\text{RA}$.
The wavemaker becomes more concentrated towards the upstream region for higher frequencies ($F \gtrsim 140$, panels \ref{wavemaker_BL} (d-g)).
This observation suggests a transition from the TS-dominated to the Orr-dominated mechanism as the frequency increases.

% Based on the quantitative and qualitative results presented above, we conclude that the mesh-free RA can accurately identify the flow instability of the non-parallel flat-plate boundary layer.

% with compact support for 

% at intermediate frequencies ($60 \lesssim F \lesssim 140$) and becomes more concentrated and towards the upstream region at high frequencies.

% At $F=20$, the overlapping region between the forcing and response modes is relatively small due to the constraint of the optimization region. 

% which signifies the activation of the Orr mechanism at high frequencies.

%  The Orr and TS mechanisms coexist and compete while both contribute to the overall energy gain of the global resolvent analysis.
% The TS and Orr mechanisms engage in a competitive interplay while contributing to the overall energy gain \citet{sipp2013characterization}.

% \newpage\clearpage

\subsection{Turbulent jet} \label{jet}

The above two examples demonstrate the capability of the proposed numerical framework for analyzing incompressible flows within the laminar regime. 
We now focus on the mean flow analysis of a turbulent iso-thermal jet at Mach number, based on the jet velocity and the far-field speed of sound, of $M=0.9$ and Reynolds number, based on the nozzle diameter and the jet velocity, of $\Re\approx10^6$. 
The large eddy simulation (LES) data, generated using the unstructured flow solver ‘Charles’, is used for analysis. Further details about the dataset can be found in \citet{bresetal_2018jfm}.
Previous studies have demonstrated that the transonic turbulent jet under consideration displays a variety of coherent features, including the well-known Kelvin–Helmholtz instabilities of the shear layer \citep{bresetal_2018jfm}, downstream non-modal Orr-type waves \citep{pickering2021optimal, schmidt2018spectral}, and trapped acoustic waves in the potential core \citep{schmidt2017wavepackets,towne2017acoustic}.

% In the following, we examine the flow instabilities around the turbulent mean flow.

% To validate the accuracy and effectiveness of the proposed mesh-free framework, 

% the lift-up mechanism at nonzero azimuthal wavenumber \citep{pickering2020lift}, and the super-directive acoustic radiation \citep{nekkanti2021modal}.

The state vector,
\begin{align}
    \vb*{q}=[\rho,u_x,u_r,u_{\theta},T]^T, \label{q}
\end{align}
comprises primitive variables: density $\rho$, temperature $T$, and cylindrical velocity components $u_x$, $u_r$ and $u_{\theta}$ in the streamwise, $x$, radial, $r$, and circumferential, $\theta$, directions, respectively. 
The fluctuating state is defined as $\vb*{{q'}}=[{\rho'},{u}_x',{u}_r',u_\theta',{T'}]^T$.
Owing to the rotational symmetry of the jet, we may decouple the governing equations and construct the compressible linear operator $\mathcal{L}$ in cylindrical coordinates, without loss of generality, for each azimuthal wavenumber $m_{\theta}$ independently. 
For a round jet, the mean azimuthal velocity component is zero.
Upon linearization of the compressible Navier-Stokes equations shown in (\ref{eqn:nse_comp}), we obtain the linearized Navier-Stokes operator around the azimuthally averaged long-time mean of the primitive state, $\overline{\vb*{q}}=[\overline{\rho},\overline{u}_x,\overline{u}_r,0,\overline{T}]^T$, such that
\begin{align}
    \pdv{}{t}\vb*{{q'}} = \mathcal{L} \vb*{{q'}}+\vb*{f}.
\end{align}
The general setup, including boundary conditions, sponge regions, and a molecular Reynolds number of $\mathrm{Re}=3\times10^4$, follows \citet{schmidt2017wavepackets} and \citet{schmidt2018spectral}.

% The configuration of sponge regions follows \citet{schmidt2017wavepackets}.
% A molecular Reynolds number of $\mathrm{Re}=3\times10^4$, as suggested in \citet{schmidt2018spectral}, is used in the computation.
% For more details, the reader is referred to \citet{schmidt2017wavepackets}.

\begin{figure}[hbt!]
\centering
\includegraphics[trim = 0mm 27mm 0mm 35mm, clip, width=.8\textwidth]{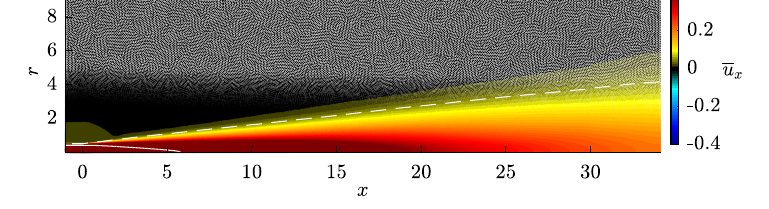}
\caption{Computational grid for jet with $N=210817\approx 1013\times 208$ nodes, colored using the mean streamwise velocity $\overline{u}_x$ at $\mathrm{Re}\approx 10^6$. The potential core (white solid) and the jet width (white dashed) are delineated as isolines corresponding to $99\%$ and $5\%$ of the jet velocity, respectively}\label{mesh_jet}
\end{figure}

The computational domain $\Omega = [-1,31.01] \times [0, 9.65]$ includes the physical domain $x,r \in [0,30] \times [0,6]$ and the surrounding sponge regions that prevent waves from being reflected.
 Local refinement is used near the nozzle with $\Delta r=0.005$ and near the jet centerline with $\Delta r=0.008$, resulting in $N = 210817$ scattered nodes for the construction of the RBF-FD-based global Jacobian $\vb*{L}$, see figure \ref{mesh_jet}. 
A structured mesh with a comparable number of $N = 185250$ was used in \citet{schmidt2018spectral}.

The resolvent gain is quantified in terms of the compressible energy norm \citep{chu1965energy} through the weight matrices
\begin{align}\label{W_chu}
    \vb*{W}_q=\vb*{W}_f\equiv \mathrm{diag}\left(\frac{\overline{T}}{\gamma \overline{\rho}M^2},\overline{\rho},\overline{\rho},\overline{\rho},\frac{\overline{\rho}}{\gamma(\gamma-1)\overline{T}M^2}\right) \otimes \mathrm{diag}(2\pi r_1\dd V_1, 2\pi r_2\dd V_2,\cdots, 2\pi r_N\dd V_N).
\end{align}
For compressible flows, the discrete weighted resolvent operator takes the form of 
\begin{align}
   \vb*{R}(\omega)= \vb*{W}_q^{\frac{1}{2}}\vb*{C}\left(\mathrm{i}\omega \vb*{I}-\vb*{L} \right)^{-1}\vb*{B}\vb*{W}_f^{-\frac{1}{2}}, 
\end{align}
where the input and output matrices, $\vb*{B}$ and $\vb*{C}$, are used to 
are used to focus the analysis exclusively on the physical domain.
As a first demonstration, we conduct the mesh-free RA for the symmetric component of the jet with $m_{\theta}=0$ in the following.

\begin{figure}[hbt!]
\centering
\includegraphics[trim = 0mm 0mm 0mm 0mm, clip, width=.55\textwidth]{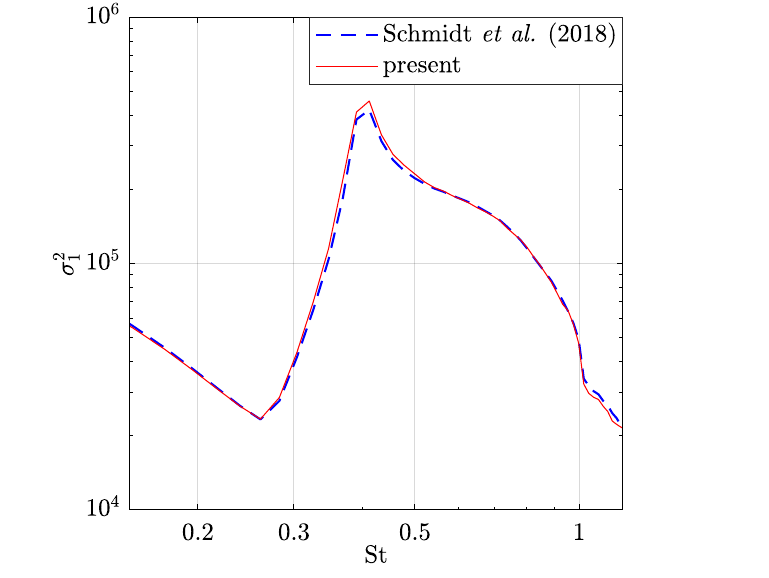}
\caption{ Leading resolvent singular value spectra for the transonic jet and $m_{\theta}=0$. Results reported by \citet{schmidt2018spectral} are shown as comparisons (blue dashed).
}\label{spectra_jet}
\end{figure}

Figure \ref{spectra_jet} compares the leading resolvent singular value spectra
to those reported by \citet{schmidt2018spectral}, obtained using a 4th-order finite difference \citep{mattsson2004summation} and a tenth-order filter for discretization.
Very good agreement is observed within the frequency range $0.15 \lesssim \mathrm{St}\lesssim 1.2$, with only a minor deviation at the peak value.
This specific frequency range has been previously identified as the regime where different physical mechanisms are active in turbulent jets, see, e.g., \citep{garnaud2013preferred,pickering2020lift,schmidt2017wavepackets,schmidt2018spectral,suzuki2006instability,tissot2017sensitivity,towne2017acoustic}.

% The observed quantitative correspondence highlights the capability of the proposed mesh-free framework to accurately predict the dominant instabilities present in turbulent flows.

% \begin{figure}[hbt!]
% \centering
% \includegraphics[trim = 0mm 12mm 0mm 0mm, clip, width=1\textwidth]{modes_comparisons_streamwise.pdf}
% \caption{
% Comparison between the PHS+poly RBF-FD discretizations (a,c,e,g,i,k) and a filtered 4th-order FDs (b,d,f,h,j,l) in terms of
% optimal response (a-f) and forcing modes (g-l)
% at three representative frequencies: (a-b, g-h) $\mathrm{St}=0.2$; (c-d, i-j) $\mathrm{St}=0.6$; (e-f, k-l) $\mathrm{St}=1$.
% The streamwise velocity component, $u_x$, is interpolated onto a stretched Cartesian mesh with contours corresponding to $\pm 0.6\|u_x\|_{\infty}$ for visualization.
% }\label{modes_jet}
% \end{figure}

\begin{figure}[hbt!]
\centering
\includegraphics[trim = 0mm 50mm 3mm 0mm, clip, width=.9\textwidth]{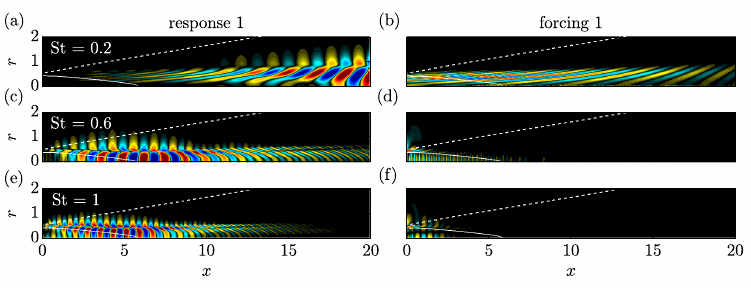}
\caption{
Streamwise velocity component of the optimal response (a,c,e) and corresponding forcing modes (b,d,f) at three representative frequencies: (a-b) $\mathrm{St}=0.2$; (c-d) $\mathrm{St}=0.6$; (e-f) $\mathrm{St}=1$.
The modes are interpolated onto a stretched Cartesian mesh with contours corresponding to $\pm 0.6\|\cdot\|_{\infty}$ for visualization. The potential core and the jet width, shown in figure \ref{mesh_jet}, are visualized for comparisons.
}\label{modes_jet}
\end{figure}

    To verify that large-scale coherent structures in the turbulent jet are accurately captured, we investigate the leading resolvent modes at three representative frequencies, as shown in figure \ref{modes_jet}.
   At $\mathrm{St}=0.2$, the optimal response exhibits a clear downstream ($x\gtrsim 10$) Orr-type wavepacket.
   The elongated structures tilted against the mean shear in the corresponding forcing are a clear indicator of the Orr mechanism associated with the optimal non-modal spatial growth \citep{schmidt2018spectral, tissot2017wave, tissot2017sensitivity}.   
   Similar flow structures have been reported in \citet{garnaud2013preferred,lesshafft2019resolvent} and \citet{pickering2021optimal}.
For higher frequencies ($\mathrm{St}\gtrsim 0.6$), the optimal response 
take the form of compact Kelvin–Helmholtz (KH) wavepackets localized upstream  
in the initial shear layer region of the jet ($x\lesssim 10$), which can be identified as the modal KH-type shear-layer instability of the turbulent mean flow \citep{gudmundsson2011instability,jordan2013wave,suzuki2006instability}.
The corresponding forcing distributions near the lip line ($r\simeq 0.5$) remain indicative of the Orr mechanism, but this time with the KH instability \citep{garnaud2013preferred,jeun2016input,qadri2017effect,semeraro2016modeling,tissot2017sensitivity}.
Within the potential core, the optimal response exhibits 
duct modes at $\mathrm{St}=0.6$ and trapped acoustic waves at $\mathrm{St}=1$. 
The presence of the latter is a general feature of resonance mechanisms between propagating waves associated with the isothermal and transonic jet, as previously described in \citep{schmidt2017wavepackets,towne2017acoustic}.
Notably, comparable patterns can be observed in the corresponding forcings, indicating the nearly self-adjoint nature of the trapped instability mechanism.
We confirmed the modal shapes are almost identical to those obtained using the numerical scheme outlined in \citet{schmidt2018spectral} (not shown).

% The above results are almost identical to those obtained using the numerical scheme outlined in \citet{schmidt2018spectral} (not shown). 

% The latter suggests the presence of resonance mechanisms between propagating waves associated with the isothermal and transonic jet \citep{schmidt2017wavepackets,towne2017acoustic}. 

% exhibit clear shear patterns aligned near the lip line ($r\simeq 0.5$), indicating that the KH instability is optimally triggered through the Orr mechanism \citep{garnaud2013preferred,jeun2016input,qadri2017effect,semeraro2016modeling,tissot2017sensitivity}.

% This agreement validates the capability of the proposed framework to accurately identify the most energetic flow structures present in turbulent jets.

% The remarkable agreement between the resolvent modes obtained in this study and those obtained using a filtered 4th-order finite difference scheme that is consistent with the one utilized in \citet{schmidt2018spectral} (not shown?) validates the capability of the proposed framework to accurately identify the most energetic flow structures present in turbulent jets.

% An upper limit on St is imposed by the numerical discretization, i.e. the capability of the differentiation scheme to resolve the smallest structures in the response and forcing fields for the given resolution.

\begin{figure}[hbt!]
\centering
\includegraphics[trim = 0mm 0mm 0mm 0mm, clip, width=.6\textwidth]{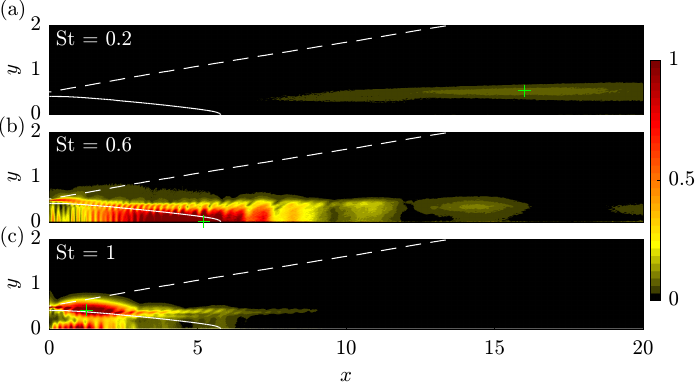}
\caption{
RA-based wavemaker $\zeta_{RA}$ for: (a) $\mathrm{St}=0.2$; (b) $\mathrm{St}=0.6$; (c) $\mathrm{St}=1$.
The region with the strongest feedback is marked as green '+'.
}\label{wavemaker_jet}
\end{figure}

Figure \ref{wavemaker_jet} shows how the RA-based wavemaker, $\zeta_{RA}$, unveils different physical mechanisms that are active in the turbulent jet. Most notably, the wavemaker region and its overall peak move upstream as the frequency increases from $\mathrm{St}=0.2$ to 1.
At $\mathrm{St}=0.2$, the optimal forcing and response modes are spatially separated (see panels \ref{modes_jet}(a,b)), resulting in a very weak wavemaker signature in the downstream self-similar region.
This result is anticipated, as the responses at low frequencies are purely triggered by the Orr mechanism without being associated with a single global mode or a local feedback mechanism that leads to the creation of waves in a localized region.
This is indicative of a non-modal convective instability.
At $\mathrm{St}=0.6$, the wavemaker peaks at the centerline at $x\simeq 5$. This is the end of the potential core, where upstream traveling acoustic modes are generated, also known as duct modes, as previously described by \citet{towne2017acoustic}.
Parallel-flow models accurately predict the occurrence of duct modes, but the wavemaker potentially reveals the location where duct modes originate, which is not predicted by the theory.
The wavemaker at $\mathrm{St}=1$ peaks within the shear layer, which is associated with the KH instability, but it also exhibits a comparable magnitude at the upstream region of the potential core, which identifies the resonance mechanisms that trigger trapped acoustic modes \citep{schmidt2017wavepackets,towne2017acoustic}.
That is, the wavemaker analysis confirms the phase-linking between downstream KH waves and upstream traveling waves at $\mathrm{St}=1$.
Note that this is not the case for $\mathrm{St}=0.6$, where the wavemaker is solely associated with the duct modes.

    % \item At $\mathrm{St}=0.2$, the wavemaker indicates the activation of the nonmodal Orr mechanism in the downstream self-similar region. The small magnitude of the wavemaker is anticipated, given the separation that exists between the forcing and response modes (see panels \ref{modes_jet}(a,b)). This indicates that the jet exhibits strong convective non-normality at low frequencies.

% This instability mechanism changes for $\mathrm{St}=0.6$ and $1$, where acoustic modes provide an upstream feedback mechanism that facilitates phase-locking between downstream and upstream waves.

% both are self-similarity?

% characteristics are linear acoustic waves

% Acoustic duct modes at $\mathrm{St}=0.6$ peak at the centerline near the end of the potential core, showing a turning point of 

% \citep{towne2017acoustic}.

% The aforementioned results provide compelling validation for the efficacy of the present mesh-free approach.

% phase-coupling

\section{Summary and discussion} \label{conclusion}

% \Red{[Summary of methods]}

In this study, a novel higher-order mesh-free framework for hydrodynamic stability analysis is developed. 
The framework has been demonstrated and validated on three benchmark problems for open flows: a canonical laminar cylinder wake, a non-parallel flat-plate Blasius boundary layer, and a transonic turbulent jet. It also provides new insights into well-known physics. 
The proposed framework incorporates PHS-type RBFs with polynomial augmentations to discretize large hydrodynamic stability matrix problems on scattered nodes with high accuracy, stability, and computational efficiency.
The resulting differentiation matrices are employed to construct the discrete linearized Navier-Stokes operator for conducting LST and RA.
We propose a set of parameters to address the trade-off between accuracy and computational efficiency for PHS+Poly RBF-FD discretizations that provide best practices.
In the context of scattered nodes, the practical implementations of various boundary conditions arising in hydrodynamic stability analysis are discussed and addressed.

The present mesh-free approach accurately predicts flow instabilities in wakes behind a 2D cylinder, including the vortex-shedding frequency, corresponding coherent structures, and structural sensitivity.
The results are in good agreement with previous studies  \citep{barkley2006linear,giannetti2007structural,marquet2008sensitivity,pier2002frequency}.
In the case of the Blasius boundary layer, the present framework accurately identifies the TS wavepackets and yields favorable quantitative and qualitative comparisons with those previously reported in the literature \citep{aakervik2008global,brandt2011effect, monokrousos2010global,sipp2013characterization}.
Furthermore, the RA-based wavemaker confirms the high non-normality of the boundary layer and the transition in dominant mechanisms from TS to Orr as the frequency increases.
When applied to the mean flow of a turbulent jet, the mesh-free RA yields results that are almost identical to those reported by \citet{schmidt2018spectral}.
The application of RA-based WM analysis identifies distinct dominant physical mechanisms operating at different frequencies. These mechanisms include the Orr mechanism in the downstream region with a very weak wavemaker signature at $\mathrm{St}=0.2$, prominent duct-like structures near the end of the potential core at $\mathrm{St}=0.6$, and the phase coupling between the KH and trapped acoustic mechanisms at $\mathrm{St}=1$.
These results closely align with the findings documented in prior literature,
e.g., \citep{jordan2013wave,lesshafft2019resolvent,schmidt2017wavepackets,schmidt2018spectral,towne2017acoustic}, and provide a new, global perspective on modal and non-modal growth in jets, as well as on the origins of these instabilities.

Building upon the pioneering work of \citet{hardy1971multiquadric}, RBF discretizations have been widely employed in canonical problems, exhibiting a continual evolution through advancements like RBF-FDs \citep{tolstykh2000using} and PHS+poly RBFs \citep{flyer2016enhancing}.
More recently, this mesh-free approach has garnered attention in computational fluid dynamics (CFD) for physical exploration or engineering applications, see, e.g., \citep{chu2023rbf,shahane2021high,shahane2022consistency, shahane2023semi, unnikrishnan2022shear}.
While RBF-FDs demonstrate the capability to construct sparse differentiation operators, the potential applications in hydrodynamic stability analysis have not been thoroughly explored.
The current work is the first demonstration of such applications.
Future extensions encompass applications involving 3D and complex geophysical or engineering fluid flows.

 % Overall, these findings suggest that the mesh-free method has great potential for advancing our understanding of complex fluid flow phenomena and for further applications in fluid mechanics research.

 % \newpage\clearpage

 \section*{CRediT authorship contribution statement}

\textbf{Tianyi Chu}: Formal analysis, Methodology, Software, Validation, Visualization, Writing – original draft. \textbf{Oliver T. Schmidt}:
Conceptualization, Funding acquisition, Methodology, Project administration, Resources, Supervision, Writing – review $\&$
editing.

\section*{Declaration of competing interest}

The authors declare the following financial interests/personal relationships which may be considered as potential competing interests: Oliver T. Schmidt reports financial support was provided by National Science Foundation under Grant No.
CBET-1953999. Tianyi Chu reports financial support was provided by National Science Foundation under Grant No. CBET1953999.

\section*{Data availability}

Data will be made available on request.
 
 \section*{Acknowledgments}

 We gratefully acknowledge support by the National Science Foundation under Grant No. CBET-1953999 (PM Ron Joslin).

\setcounter{equation}{0}
\renewcommand{\theequation}{{\rm A}.\arabic{equation}}

\appendix

\section{Governing equations for compressible flows} \label{Governing_compressible}

The compressible Navier-Stokes equations govern the motion of a general, compressible Newtonian fluid,
\begin{equation}
    \begin{aligned}
    \label{eqn:nse_comp}
  \frac{\partial \rho}{\partial t}&  = -\nabla\cdot \rho \mathbf{u}, \\[3pt]
  \frac{\partial \rho \mathbf{u}}{\partial t}& =
  -\frac{1}{2}\nabla\cdot(\mathbf{u}:\rho\mathbf{u} + \rho\mathbf{u}:\mathbf{u}) -
  \nabla p + \frac{1}{\mathrm{Re}}\nabla\cdot \vb*{\tau}, \\[3pt]
  \frac{\partial \rho e}{\partial t}& = -\nabla\cdot \rho e \mathbf{u} +
  \frac{1}{(\gamma-1)\mathrm{Re}\mathrm{Pr}\mathrm{Ma}^2_\infty}\nabla\cdot k \nabla T - \nabla\cdot p \mathbf{u} +
  \frac{1}{\mathrm{Re}}\nabla\cdot\vb*{\tau}\mathbf{u},
    \end{aligned}
\end{equation}
where, $e$ is the total energy. For a Newtonian fluid, the viscous stress tensor is $\vb*{\tau} = \mu \left( \nabla\mathbf{u} +  \nabla\mathbf{u}^T \right) -\frac{2}{3}\mu(\nabla\cdot \mathbf{u})\mathrm{I}$. All flow quantities are non-dimensionalized by their dimensional free-stream values, denoted by $(\cdot)_{\infty}^*$, and the coordinates by the jet diameter $D$. The dimensionless Reynolds number $\mathrm{Re}={\rho_{\infty}^*u_{\infty}^*D}/{\mu_{\infty}^*}$, Prandtl number $\mathrm{Pr}={\mathrm{c_{p}^*}\mu_{\infty}^*}/{k_{\infty}^*}$, and Mach number $M={u_{\infty}^*}/{a_{\infty}^*}$ then fully describe the flow. Here, $\mu_{\infty}^*, k_{\infty}^*, c_{p}^*, \gamma, a_{\infty}^*$ are the free-stream values of the dynamic viscosity, heat conductivity, heat capacity at constant pressure, heat capacity ratio, and speed of sound, respectively. Closure of the equations is achieved under the assumption of an ideal gas, and using Sutherland's law to compute the dynamic viscosity from the local temperature.

% \section{Boundary layer}

 \end{document}